\documentclass[aps,prx,onecolumn,floatfix,superscriptaddress,showpacs,amsmath,amssymb]{revtex4-2}
\usepackage{graphicx}
\usepackage[dvipsnames]{xcolor}
\usepackage{bm}
\usepackage{hyperref}
\usepackage[capitalize]{cleveref}
\crefrangeformat{equation}{Eqs.~(#3#1#4)-(#5#2#6)}

\newcommand{\bsf}[1]{\textsf{\textbf{#1}}}

\def\Tq{\bsf{T}_{\qq}}

\def \uu  {{\bm u}}
\def \qq  {{\bm q}}

\def \Rey  {\mbox{Re}}

\usepackage[normalem]{ulem}
\newcommand{\beq}{\begin{equation}}
\newcommand{\eeq}{\end{equation}}

\newcommand{\ADD}[1]{{#1}}

\newcommand{\REM}[1]{{}}

\newcommand{\Eq}[1]{Eq.~(\ref{#1})}
\newcommand{\Fig}[1]{Fig.~(\ref{#1})}
\newcommand{\subfig}[2]{Fig.~(\ref{#1}#2)}

\begin{document}

\title{Inertia drives a flocking phase transition in viscous active fluids}

\author{Rayan Chatterjee}
\affiliation{TIFR Centre for Interdisciplinary Sciences, Tata Institute of Fundamental Research,  Gopanpally, Hyderabad 500046, India}
\author{Navdeep Rana} 
\affiliation{TIFR Centre for Interdisciplinary Sciences, Tata Institute of Fundamental Research,  Gopanpally, Hyderabad 500046, India}
\author{R. Aditi Simha}\thanks{deceased}
\affiliation{Department of Physics, Indian Institute Of Technology Madras, Chennai 600 036, India}
\author{Prasad Perlekar}
\affiliation{TIFR Centre for Interdisciplinary Sciences, Tata Institute of Fundamental Research,  Gopanpally, Hyderabad 500046, India}
\author{Sriram Ramaswamy}\thanks{Adjunct Professor, Tata Institute of Fundamental Research, Feb 2017 - Jan 2020}
\affiliation{Centre for Condensed Matter Theory, Department of Physics, Indian Institute of Science, Bangalore 560 012, India}
\begin{abstract} 

How fast must an oriented collection of extensile swimmers swim to escape the instability of viscous active suspensions?
We show that the answer lies in the dimensionless combination $R=\rho v_0^2/2\sigma_a$, where $\rho$ is the suspension mass density, $v_0$ the swim speed and $\sigma_a$ the active stress. 
Linear stability analysis shows that for small $R$ disturbances grow at a rate linear in their wavenumber $q$, and that the dominant instability mode involves twist.
The resulting steady state in our numerical studies is isotropic hedgehog-defect turbulence. Past a first threshold $R$ of order unity we find a slower growth rate, of $O(q^2)$; the numerically observed steady state is {\it phase-turbulent}: noisy but {\it aligned} on average. We present numerical evidence in three and two dimensions that this inertia driven flocking transition is continuous, with a correlation length that grows on approaching the transition. 
For much larger $R$ we find an aligned state linearly stable to perturbations at all $q$. Our predictions should be testable in suspensions of mesoscale swimmers [D Klotsa, Soft Matter \textbf{15}, 8946 (2019)].
\end{abstract}

\maketitle

\section{Introduction} \label{sec:intro} 
The theory of active matter~\cite{mechstatact,rmp,tonertu_sr,sr_actfluid,prost_actgel,julicher_cyto,jul18,mar16,Gompper_2020} -- systems whose constituents convert a sustained supply of fuel into movement -- is the framework of choice for understanding the collective behaviour of motile particles. Like condensed matter in general, active systems display many types of order and operate in a variety of dynamical regimes. Our interest in this paper is in groups of motile organisms in a bulk fluid medium, spontaneously organized into a flock in which their tail-to-head vectors on average point in a common direction. In now-standard terminology \cite{rmp}, we consider \textit{polar, wet} active matter, described by a \textit{vector} order parameter characterizing the degree and direction of common orientation and movement. 

In the world of Stokesian hydrodynamics, where inertia is absent and viscosity holds sway, an ordered flock in bulk fluid is impossible~\cite{SIMSR2002}. Indeed this limit has shaped the defining image of active suspensions as inescapably unstable, dissolving via spontaneous flow~\cite{SIMSR2002,Voit1} and defect proliferation~\cite{sagues,silberzan,SaintillanARFM,giomi_polarPRL,giomi2,giomi3,URZAY2017,gold_lauga,kruserotatingdefect,marchettidefect,elgeti,julia_natcom,juliasumesh1,giomi4,sanchez,dogic} into a kind of turbulence~\cite{shell2007,wol2008,actnem18,gold_lauga,Dombrowski,Ishikawa1,Sokolov,wensink1,dunkel1, bratanov,needleman_stormy}. We remind the reader of the instability mechanism \cite{SIMSR2002,mechstatact,rmp}: an aligned state of active particles is a state of uniform uniaxial stress; perturbing this state -- through bend or splay respectively for ``pusher'' or ``puller'' particles -- creates spatially varying stresses; force balance requires that these are accompanied by flow; this flow rotates the alignment further in the direction of the perturbation. Note that this description refers neither to the directed motion of the particles nor to acceleration. Stresses and flows are uniaxial but \textit{apolar}, that is, fore-aft symmetric, and flow responds \textit{instantaneously} to active stress in the Stokesian approximation. Improved descriptions including polar order alone \cite{SIMSR2002,giomi_polarPRL,giomi2,tjhung08}, or inertia alone \cite{SIMSR2002}, do not mitigate the instability.

Stable flocks in bulk fluid are of course widely observed in the form of fish schools \cite{partridge1982structure,weihs1973hydromechanics}, which are very polar and very far from Stokesian. We do not venture into the regime of schooling at high Reynolds number, governed by purely inviscid hydrodynamic interactions \cite{Shelley_flapping,Kanso_PhysRevLett.120.198101}, but consider weak inertial effects, which are know to alter significantly the viscous hydrodynamic interaction between slow swimmers~\cite{ardekani1,ardekani2}. A recent Perspective \cite{klotsa_persp} makes a persuasive case for the study of active fluids with small but non-negligible inertia, the regime we explore here. A result from \cite{SIMSR2002} is relevant in this context: a linearized treatment retaining only acceleration and active stresses finds a parameter domain in which flocks in fluid are neutrally stable to first order in wavenumber, with a wavelike dynamic response. Interestingly, such waves of bend excitations have an analog in models without momentum conservation, i.e., ``dry'' flocking models, when rotational inertia is taken into account \cite{attanasi2014information,cavagna2015silent,Chate_memory,dadhichi2020nonmutual}; the coupled dynamics of classical spin angular momentum and orientation in such models is formally similar to that of hydrodynamic vorticity and orientation in \cite{SIMSR2002} and the present work. Staying close to viscous hydrodynamics and the force-dipole picture of swimmers, we ask: can inertia and polar order together defeat the Stokesian instability of flocks?

\subsection{Summary of results} \label{subsec:results}
\begin{figure*}[!t]
    \centering
	\includegraphics[width=0.65\linewidth]{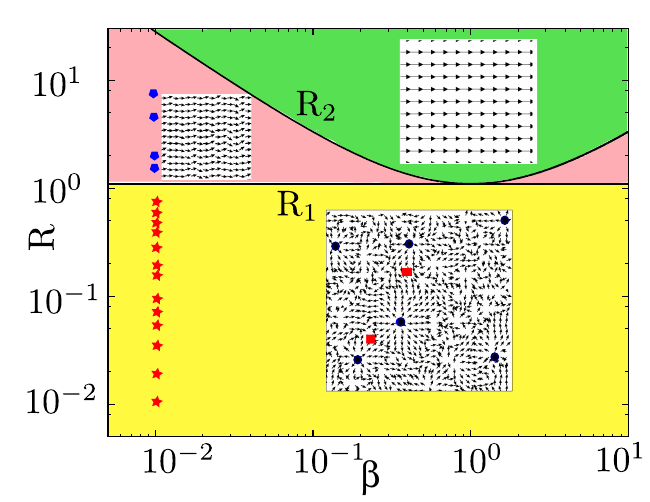}
    \caption{ \label{fig:Rvsbeta}
        $R$-$\beta$ phase diagram of a polar active suspension. The lines at $R=R_1$ and $R=R_2$, obtained from linear
            stability analysis, mark the phase boundaries. For $R < R_1$, an aligned state is $O(q)$ unstable which leads to
            statistically stationary defect turbulence.   In general, a parameter range ($R_1 < R< R_2)$ where perturbations
            with small wavenumber $q$ grow at a rate of $O(q^2)$ intervenes between the stable regime ($R>R_2$) and the highly
            unstable regime of $O(q)$ growth, but is squeezed out of existence if $\beta =1$, that is, orientation and vorticity
            have identical diffusivities.  The red stars and blue pentagons mark the ($R$, $\beta$) values used in our DNS.
            (Insets) Order parameter vector field for defect turbulence ($R<R_1$) with asters (blue dot)
        and saddles (red square), phase turbulence with orientational order ($R_1<R<R_2$), and a quiescent linearly stable state of complete alignment ($R>R_2$).}
\end{figure*}

In this article, focusing on \textit{extensile} or ``pusher'' \cite{Lauga_2009} suspensions, we show that the introduction of inertia qualitatively alters our understanding of the viscous hydrodynamics of polar active matter, that is, flocking in fluids. Here are our main results. We show that speed matters: the dimensionless combination $R \equiv \rho v_0^2/2 \sigma_a$, where $\rho$ is the suspension mass density, $\sigma_a$ the scale of active stress, and $v_0$ the self-advection speed, governs the stability of active suspensions. Flocks in fluid are \textit{stable} for large $R$, and their inexorable Stokesian instability \cite{SIMSR2002} is the $R=0$ limit of a far richer picture, Fig. \ref{fig:Rvsbeta}. 
For small $R$ 
perturbations about the aligned state grow at a rate $\propto q$ for wavenumber $q\to 0$ while for moderate $R$ 
the linear instability persists but with a growth rate $\propto q^2$. Crucially, direct numerical simulations of the hydrodynamic equations reveal that the two regimes 
correspond to qualitatively distinct statistical steady states separated by a nonequilibrium phase transition. The small-$R$ regime is isotropic hedgehog-defect turbulence while that at moderate $R$ 
is a phase-turbulent \cite{chatemanneville,Shraiman_phs_trb,Chate_phs_trb,Chate_phs_dgm} 
but \textit{ordered} flock. Our numerical results suggest a continuous order-parameter onset and a growing correlation length upon approaching the transition.

This paper is organized as follows. In \S~\ref{sec:framework} we present the equations of hydrodynamics for polar active suspensions and investigate the linear stability of the uniaxially ordered state. \S~\ref{sec:numerics} describes our numerical studies, the flocking transition from defect to phase turbulence, and the properties of the turbulent states. We close in \S~\ref{sec:concl} with a summary, suggestions for experiment, and open questions.
 
\section{ \label{sec:framework} Governing equations and stability analysis} 
\subsection{\label{subsec:activehydro} Hydrodynamics of active suspensions}  

\ADD{We begin by constructing, from general principles, the hydrodynamic equations of motion for a flock in fluid. We do not employ the language of forces and fluxes or display the dependence of ``active'' coefficients on a maintained chemical driving force \cite{rmp,prost_actgel,dadhichi2018origins}. We adopt the general, symmetry-based approach of ~\cite{SIMSR2002} but our treatment is self-contained and does not presuppose familiarity with that work. The reader will see the results of \cite{SIMSR2002} emerge as a limiting case in section \ref{subsub:largeq}.} We emphasize that our equations constitute a general effective description on length-scales much larger than a swimmer (as we shall call our self-propelled particles hereafter). They contain parameters such as viscosity and elastic constants; these are phenomenological coefficients in our coarse-grained description of this internally driven system, and are named based on the form of the terms they govern. Their values cannot in principle be estimated from a near-equilibrium hydrodynamic theory of the suspension. For example, we would imagine the viscosity in our equations receives ``eddy'' contributions from flows on scales of a few swimmers, and we expect that the elastic constants encoding the aligning tendency are at least partly behavioural rather than mechanical. Provided interactions are local in space and time, these features do not limit the validity of our approach, which depends only on conservation laws and symmetries. 

For a steady state such as a flock, which spontaneously breaks a continuous invariance, the slow or hydrodynamic variables \cite{MPP} are the local densities of conserved quantities and the broken-symmetry or Nambu-Goldstone \cite{nambu1961dynamical,goldstone1961field} fields. At a continuous transition to such an ordered state, the amplitude of the order parameter is an additional slow variable. In the absence of reproduction, death, and external forces, the conserved quantities are the total number of swimmers, the total amount of fluid, and the total momentum of \textit{swimmers plus fluid}. Energy conservation does not play a role as each swimmers is endowed with a built-in power source. The slow variables corresponding to these conservation laws are then the densities $\rho$ and $\bm{g} = \rho \bm{u}$ of mass and momentum of swimmers plus fluid (defining the suspension velocity field $\bm{u}$), and the number density $c$ of swimmers. The broken-symmetry modes and the magnitude of order are jointly contained in the polar order parameter field $\bm{p}$, which is the local average of the orientation unit vectors of the particles \cite{SIMSR2002}. \ADD{It is interesting to note that an \textit{equilibrium} liquid crystal with macroscopic vectorial, i.e., polar, order has only very recently been discovered \cite{chen2020first}.}

As the order parameter is a space vector, our description is invariant under the joint inversion of $\bm{p}$ and the spatial coordinate $\bm{r}$, but not $\bm{p}$ alone. The absence of $\bm{p} \to -\bm{p}$ symmetry is central to our narrative. We therefore include at this stage, at leading order in gradients, all terms that break this symmetry in the equations of motion \cite{maitra2014activating,maitra_polar_confined} (see also \cite{MCM_TBL_bookchap_Lenz,kung,rmp,giomi_polarPRL,giomi2,tjhung2012spontaneous}), although we will shortly pass to a more economical description. The equations read 
\begin{widetext}
\begin{eqnarray} 
\label{eq:vel}
\rho (\partial_t  \uu +\uu \cdot \nabla \uu) &=& -\nabla P + \mu \nabla^2 \uu + \nabla \cdot ({\bm \Sigma}^a+{\bm \Sigma}^r), \\
\label{eq:dir}
\partial_t\bm{p} + (\uu + v_0{\bm p})\cdot\nabla {\bm p}&=& \lambda {\bm S}\cdot\bm{p}+{\bm \Omega}\cdot{\bm p} + \Gamma {\bm h} +\ell \nabla^2 {\uu}, ~\rm{and}\\
\label{eq:conc}
\partial_t c+\nabla\cdot\left[(\bm{u}+v_1\bm{p})c\right]&=& 0.
\end{eqnarray}
\end{widetext}

\ADD{Equation \eqref{eq:conc} expresses number conservation; the active particles self-propel with velocity $v_1 \bm{p}$ in the frame of the suspension, and hence $\bm{u} + v_1 \bm{p}$ in the laboratory frame. In \eqref{eq:dir}, the polar order parameter $\bm{p}$ is carried by the hydrodynamic velocity $\bm{u}$ and by its self-advection $v_0$ [not related by any symmetry \cite{toner1998flocks,dadhichi2020nonmutual} to $v_1$ in \eqref{eq:conc}], ${\bm S}$ and ${\bm \Omega}$ are the symmetric and antisymmetric parts of the velocity gradient tensor $\nabla \uu$ which couple orientation to flow as in ordinary nematic liquid crystals \cite{degp}, $\Gamma$ is the kinetic coefficient governing relaxation in the local molecular field $\bm{h}$ to be discussed below, and $\ell$ is the polar flow coupling at leading order in a gradient expansion. The $v_0$, $v_1$ and $\ell$ terms are polar: their presence implies that the equations are not invariant under $\bm{p} \to - \bm{p}$. In \eqref{eq:vel}, the hydrodynamic pressure $P$ enforces incompressibility $\nabla\cdot \uu=0$.}  
\begin{equation} 
\label{activestress}
{\bm \Sigma}^a \equiv -\sigma_a \bm{p}\bm{p} - \gamma_a(\nabla \bm{p} + \nabla \bm{p}^T)
\end{equation}
\ADD{is the intrinsic stress associated with swimming activity, which we display up to first subleading order in gradients. In \eqref{activestress} $\sigma_a$ is the force-dipole density~\cite{SIMSR2002,hatwalne2004rheology,rmp}. In microscopic terms, the forces exerted by a swimmer and the ambient fluid on each other add to zero, so the associated force density has zero monopole moment. The minimal model for a swimmer is thus a point dipolar force density. A collection of such swimmers, each with dipole strength $W$, local concentration $c$, and mean local alignment given by the polar order parameter $\bm{p}$ can readily be seen \cite{SIMSR2002,hatwalne2004rheology,rmp} to have force density $-W\nabla \cdot (c \bm{p}\bm{p})$. $W>0$ and $W<0$ correspond respectively to extensile swimmers, that push fluid back with their tails and move forward, and contractile swimmers, that advance by pulling fluid toward themselves from the front. Thus, in \eqref{activestress}, $\sigma_a = W c$. The {\it polar} contribution to the active stress, given at leading order in a gradient expansion by the $\gamma_a$ term in \eqref{activestress}, arises \cite{MCM_TBL_bookchap_Lenz,hatwalne2004rheology} if the force dipole on each particle is displaced with respect to the center of drag of the particle, as it must be to achieve locomotion. From the foregoing it is plausible that $\gamma_a$ should be proportional to $\sigma_a$, with the proportionality factor being a length that measures the fore-aft asymmetry of the active particles. In principle all parameters in our equations should be functions of
the local concentration $c$. Phenomenologically, $\sigma_a, \, \gamma_a$ are tied to the presence of active particles and should therefore be proportional to $c$ for $c \to 0$. In the simple microscopic picture discussed above, if $W$ is treated as an intrinsic single-particle property and therefore independent of $c$, the proportionality is exact. The contribution}
\begin{equation} 
\label{passivestress}
{\bm \Sigma}^r = {1 - \lambda \over 2}\bm{p}\bm{h}  - {1 + \lambda \over 2} \bm{h}\bm{p} 
-\ell (\nabla \bm{h} + \nabla \bm{h}^T) 
\end{equation}
\ADD{is the reversible thermodynamic stress for an equilibrium polar liquid crystal. The expression \eqref{passivestress} extends the form found in ~\cite{kruse1,rmp} to include the leading-order polar $\ell$ term, which is the Onsager counterpart of the polar flow-coupling term $\ell \nabla^2 \bm{u}$ in \eqref{eq:dir}, discussed in \cite{maitra2014activating} and \cite{maitra_polar_confined}.} $\bm{h} = - \delta F / \delta {\bm p}$ is the molecular field conjugate to ${\bm p}$, derived from a free-energy functional 
\begin{equation} \label{freeNRG}
F = \int d^3r \left[\frac{1}{4}({\bm p}\cdot{\bm p} - 1)^2+ \frac{K}{2} (\nabla {\bm p})^2 - E {\bm  p} \cdot \nabla c\right]
\end{equation}
favouring a ${\bm p}$-field of uniform magnitude~\cite{concdepfoot} which we have rescaled to unity. A single Frank constant \cite{Frank,zocher,oseen} $K$  penalizes gradients in ${\bm p}$, and $E$ promotes alignment of ${\bm p}$ up or down gradients of $c$, according to its sign. $\mu$ is the shear viscosity of the suspension, and $\Gamma$, the collective rotational mobility for the relaxation of the polar order parameter field, is expected to be of order $1/\mu$. $\lambda$ is the nematic flow-alignment parameter~\cite{forster,stark} and $\ell$, with units of length, governs the lowest-order {\it polar} flow-coupling term \cite{maitra2014activating,maitra_polar_confined}. If our equations were derived from a microscopic model of particles in a fluid, we expect that both $\ell$ and the length $\gamma_a/\sigma_a$ would be related to a fore-aft asymmetry in the dimensions of the active particles.
	
\subsubsection{\ADD{Essential and incidental polar contributions}}\label{subsub:simplepolar} 
Equations \eqref{eq:vel}-\eqref{eq:conc} are endowed with a surfeit of parameters originating in the polar character of our system -- the speeds $v_0$ and $v_1$ at which the orientation advects itself and the concentration respectively, the polar active stress coefficient $\gamma_a$ and the passive polar flow-coupling length-scale $\ell$.
\ADD{In the work \cite{SIMSR2002} that initiated the study of the hydrodynamics of active liquid crystals, the polar character of the order parameter of a flock played an important role, combining with inertia to yield a propagative mode-structure. As this was a leading-order feature in a gradient expansion, neither $\gamma_a$ nor $\ell$, which enter at next-to-leading order relative to $\sigma_a$ in \eqref{activestress} and $1 \pm \lambda$ in \eqref{passivestress} and \eqref{eq:dir} respectively, were considered, and polar effects thus entered \cite{SIMSR2002} only through $v_0$ and $v_1$. The instability of active liquid crystals in the Stokesian domain -- the other major finding of \cite{SIMSR2002} -- commanded much greater interest in the field thanks to its connection to experimental realizations in cellular and microbial settings. An analysis that ignores polarity altogether and works only with the \textit{axis} of orientation offers a satisfactory conceptual understanding of that instability \cite{rmp,actnem18}, though interesting complexities arise \cite{giomi_polarPRL} in a Stokesian setting through the polar parameters $\gamma_a$, $v_0$ and $v_1$. A final remark in this context is that polarity will assert its presence in any formulation in terms of a vector order parameter $\bm{p}$, even if the equations of motion are invariant under $\bm{p} \to - \bm{p}$, through the nature of topological defects \cite{kruserotatingdefect,rana2020}.}
	
\ADD{For the purposes of the present work what matters is that the self-advection speed $v_0$ plays a distinct -- and crucial -- role. The other polar parameters contribute in an incidental manner.} First, we are concerned here only with the extensile case $\sigma_a>0$, for which the instability mode is bend, which decouples from concentration in the linear theory. In what follows we therefore ignore the concentration field, and hence $v_1$ drops out of our analysis. Next, as we show in the Appendix, it is only through $v_0$ -- in the form $\rho v_0^2$ and its competition with $\sigma_a$ -- that the stabilizing effects of inertia enter our treatment. $\gamma_a$ and $\ell$ leave unaltered both the coefficient of the $O(q)$ contribution to the mode frequency and the parameter value at which the instability growth rate changes from $O(q)$ to $O(q^2)$. They simply shift the coefficients of the $O(q^2)$ piece of the mode frequency by amounts of relative order unity. We therefore work with an economical description in which $\gamma_a$ and $\ell$ are zero and polar effects enter only through $v_0$ and, of course, the nature of the allowed topological defects. Crucially, $v_0$ and $\sigma_a$ are independent quantities in our coarse-grained treatment, a point we will return to later in the paper.

\subsection{Linear Stability analysis} \label{subsec:linstab}
Defining the ordering direction to be $\hat{\bm x}$ and directions in the $yz$ plane as $\perp$, we have investigated the stability of a uniform ordered flock ($c=c_0, {\bm u}={\bm 0}$, and ${\bm p}=  \hat{\bm x}$, which is a stationary solution of Eqs.\eqref{eq:vel} - \eqref{eq:conc} to small perturbations  ($\delta {\bm u}_{\perp},\delta {\bm p}_{\perp},\delta c$), where the presence of only the $\perp$ components is a result of incompressibility and the ``fast'' nature of $p_x$. We present here the results for the case where the concentration field $c$ is removed from the analysis. This is sufficient for our purposes, because $c$ does not participate significantly in the linear instabilities of relevance, as we now argue. Taking the curl with respect to $\nabla_{\perp}$ eliminates $c$ from the $\perp$ component of \Eq{eq:dir}. A similar curl removes it from the $\perp$ component of \Eq{eq:vel} as well. 
Thus concentration does not participate in the linear dynamics of the twist-bend mode \cite{SIMSR2002}. The 3-dimensional instability of extensile active fluids is known, numerically, to be twist-dominated \cite{shen}, an observation for which our linear stability analysis below provides the natural explanation. A description without a concentration field should thus be a reasonable guide to instabilities and active turbulence in our system. It is important to note that the neglect of the concentration field in our treatment does \textit{not} amount to an incompressibility constraint on the polar order parameter field. A formal connection between the complete equations and those without a concentration field can be achieved by introducing birth and death of particles so that $c$ becomes ``fast'' \cite{malthus1} and can be eliminated in favour of the slow variables $\bm{p}_{\perp}$ and $\bm{u}_{\perp}$, with at most a finite shift in parameter values in the equations for the slow variables. Changes in the linear stability analysis upon inclusion of the concentration are quantitative, not qualitative, and can be found in the Appendix. Defining the projector 
\begin{equation}
\label{eq:transproj}
\Tq \equiv \bsf{I} - \hat{\bm q} \hat{\bm q}
\end{equation}
transverse to $\bm{q}$ and linearizing Eqs.~\eqref{eq:vel} and~\eqref{eq:dir} about the ordered state we find
	\begin{eqnarray}
		\label{eq:lu_nc}
	\left(\rho\partial_t + \mu q^2\right)\delta \uu_{\perp \qq}
	&=&  - i\Tq \cdot  \left[\left(\sigma_a +{\lambda-1 \over 2} K q^2\right) \hat{\bm x} \qq_{\perp} + q_x \left(\sigma_a +{\lambda + 1 \over 2}K  q^2\right)\bsf{I}\right] \cdot\delta \bm{p_{\perp \qq}}\\
	\label{eq:lp_nc}	
	\partial_t  \delta \bm{p}_{\perp \qq}
	&=&  +i\left({\lambda+1 \over 2}q_x \bsf{I}  
	-{\lambda-1\over 2} {\qq_{\perp}\qq_{\perp} \over q_x}\right)\cdot\delta \bm{u}_{\perp \qq}  - \left(iv_0q_x + \Gamma K q^2\right)\delta \bm{p}_{\perp \qq}.
	\end{eqnarray} 
As in \cite{SIMSR2002}, the divergence and curl of Eqs.~\eqref{eq:lu_nc} and \eqref{eq:lp_nc} describe respectively the dynamics of splay and twist, with an admixture of bend in each case for $q_x \neq 0$. Defining $\phi$ to be the angle between the wavevector $\bm{q}$ and the alignment ($\hat{\bm x}$) direction, the resulting dispersion relations for the frequency $\omega$, valid for all $\bm{q}$, for modes of the form $e^{i(\bm{q}\cdot {\bm{r}}-\omega t)}$, are 
\begin{eqnarray}
\label{eq:omvsq}
\omega = \omega^s_{\pm}= {1 \over 2} v_0 q\cos\phi-i{\mu_{+} \over 2 \rho} q^2 \pm \left(\sigma_a \over 2\rho\right)^{1/2} \left[ A(\phi) q^2+ iB(\phi)q^3+G(\phi)q^4 \right]^{1/2}
\end{eqnarray}
for the splay-bend modes 
and 
\begin{eqnarray}
\label{eq:omvsq34}
\omega^t_{\pm}= {1 \over 2}v_0 q\cos\phi- i{\mu_+ \over 2 \rho} q^2 \pm \left(\sigma_a \over 2\rho\right)^{1/2} \left[A(0)\cos^2\phi q^2+ iB(0)\cos\phi q^3+\bar{G}(\phi)q^4 \right]^{1/2}
\end{eqnarray}
for the twist-bend modes. In Eqs.~(\ref{eq:omvsq}) and~(\ref{eq:omvsq34}) we have defined $A(\phi)\equiv R\cos^2\phi- \cos 2\phi(1+\lambda\cos 2\phi)$, $B(\phi)\equiv (v_0\mu_-/\sigma_a) \cos\phi$, $G(\phi)=-(\mu_-^2/2 \rho \sigma_a) + (K/2\sigma_a)(1+\lambda\cos 2\phi)^2$, and $\bar{G}(\phi)\equiv - (\mu_-^2/2 \rho \sigma_a) +  (K/2\sigma_a)(1+\lambda)^2\cos^2\phi$ \cite{GGbarfoot} with 
\begin{equation}
\label{Rdef}
R \equiv \rho v_0^2/2\sigma_a, 
\end{equation}
and $\mu_{\pm}\equiv{\mu} (1 \pm \beta)$ where $\beta \equiv \Gamma K \rho/\mu$ should be of the same order as $\alpha \equiv K \rho/\mu^2$ because the mobility $\Gamma \sim 1/\mu$. For conventional liquid crystals $\alpha, \, \beta \ll 1$. 
 
When $R=0$, the extensile ($\sigma_a > 0$) systems of interest here present a bend instability [see Eqs. \eqref{eq:omvsq} and \eqref{eq:omvsq34}] with invasion speed $\sqrt{\sigma_a/\rho}$. For $v_0 > 0$ disturbances can outrun this invasive growth. The dimensionless combination $R$ describes this competition. Note that the contribution of $R$ vanishes for pure splay, Eq.~(\ref{eq:omvsq}) at $\phi = \pi/2$, so motility cannot stabilize contractile ($\sigma_a < 0$) flocks in fluid.

\subsubsection{Small-$q$ behaviour: the $O(q)$ and $O(q^2)$ instabilities} \label{subsub:smallq} Let us first examine the small-$q$ behaviour.
Expanding Eqs.~(\ref{eq:omvsq}) and~(\ref{eq:omvsq34}) up to order $q^2$ we then find  
 \begin{equation}
\label{eq:smallq12}
\omega =  \omega^s_{\pm}
={q \over 2} \left\{  v_0 \cos\phi \pm   \left[{2\sigma_a \over \rho} A(\phi)\right]^{1/2} \right\}
-{i \over 2}{\mu \over \rho} q^2\left\{1 + \beta \mp  (1 - \beta)  \left[
R \cos^2 \phi \over A(\phi)\right]^{1/2}\right\}
\end{equation}
for the splay-bend modes and
\begin{equation}
\label{eq:smallq34}
\omega =  \omega^t_{\pm} = {q \over 2} \cos\phi\left\{  v_0  \pm   \left[{2\sigma_a \over\rho}A(0)\right]^{1/2}\right\}
-{i \over 2}{\mu \over \rho} q^2  \left\{1 + \beta  \mp (1 - \beta) \left[{R \over A(0)} \right]^{1/2}\right\}
\end{equation}
for the twist-bend modes. Here $A(0) = A(\phi=0) = R -(1+\lambda)$. One note of caution: the small-$q$ expansion that led to Eq.~(\ref{eq:smallq34}) assumes $v_0q \cos \phi > q^2 \mu/\rho$, which means that it does not apply for $\phi = \pi/2$, i.e., pure twist. It does however hold for any $\phi \in [0, \pi/2)$ but the closer $\phi$ is to $\pi/2$ the smaller $q$ must be for the result to apply.

Two of our main results now follow. If $R < 1+\lambda$, Eq.~(\ref{eq:smallq34}) signals a bend instability with small-$q$ growth rate $\sim q$. This was discussed in the strictly apolar case $v_0=0$ in \cite{SIMSR2002}, and can be viewed as the small-$q$ extension of the Stokesian bend instability \cite{SIMSR2002}. However, if $R > 1+\lambda$, so that the $O(q)$ instability is averted,  $0 < 1 - (1+\lambda)/R < 1$. If $R$ is not too large, this means the coefficient of $iq^2$ in Eqs.~\eqref{eq:smallq12} and ~\eqref{eq:smallq34} is positive, signalling a small-$q$ instability with \textit{diffusive} growth. This $O(q^2)$ instability exists for $R$ between $R_1 = 1 + \lambda$ and 
\begin{equation}
\label{R2}
R_2 = {\mu_+^2 \over \mu_+^2 - \mu_-^2}R_1 = {1 + \lambda \over 4 \beta} (1 + \beta)^2.
\end{equation}  
For $R>R_2$ the flock is linearly stable. If $\beta \ll 1$ as in molecular systems, $R_2 \gg R_1$, and the $O(q^2)$ instability occupies a large range of $R$. In the $\beta = 0$ limit the uniformly ordered flock is \textit{always} linearly unstable, with small-$q$ growth rate $\sim q$ for $R < 1 + \lambda$ and $\sim q^2$ for $R > 1+\lambda$. Fig. \ref{fig:Rvsbeta} summarizes the small-$q$ stability behaviour.

Note that the $O(q^2)$ instability can be eliminated in the special case $\beta = 1$, i.e., $\mu/\rho = \Gamma K$. Noting that $\Gamma$ should be roughly $1/\mu$, this condition implies $K= \mu^2/\rho$, an interesting condition that equates a Frank constant (which, recall, has units of force in three dimensions) to Purcell's intrinsic force scale \cite{Purcell} $\mu^2/\rho$ for three-dimensional viscous fluids. As we remarked above, $\beta$ in molecular or colloidal systems is about $10^{-4}$ \cite{Orsay_nemdyn,degp}, so requiring it to be order unity amounts to insisting that the swimmers have an exceptionally strong aligning interaction. This possibility cannot be ruled out a priori as alignment in living systems is likely to be active and behavioural, not a passive mechanical torque. 

\subsubsection{Large-$q$ dynamics and the Stokesian limit} \label{subsub:largeq} 
\ADD{Having established the general linearized behaviour of active extensile liquid crystals in the true hydrodynamic regime of small wavenumbers, we turn our attention to large wavenumbers. Two length scales are important here -- }
\begin{equation}
\label{eq:ell}
\ell_v \equiv \mu /v_0 \rho \, \, \mbox{and} \,\, 
\ell_\sigma \equiv \mu /\sqrt{\rho \sigma_a} = R^{1/2} \ell_v
\end{equation}
below which viscosity overwhelms the inertial effects of self-advection 
and 
\begin{equation}
\label{eq:ellK}
\ell_K\equiv \sqrt{K/\sigma_a}
\end{equation}
below which Frank elasticity dominates active stresses. \ADD{For molecular or colloidal systems, for which, as we remarked earlier, $\alpha$ is exceedingly small \cite{degp,Orsay_nemdyn}, ${\ell_K/\ell_v} = \sqrt{\alpha R}$ should be small too except in the unlikely condition of ultra-high self-advection speeds. The wavenumber range $\mbox{max}(\ell_v^{-1},\ell_\sigma^{-1}) \ll q \ll \ell_K^{-1}$ should be substantial. }
Expanding Eqs.~(\ref{eq:omvsq}) and (\ref{eq:omvsq34}) for $q  \gg \mbox{max}(\ell_v^{-1},\ell_\sigma^{-1})$, we find, to leading order in $\alpha$ and $\beta$, that the splay-bend mode that goes unstable at small $R$ has the form
\begin{widetext}
\begin{eqnarray}
\label{splay_largeq}
\omega^s &=& -i {\sigma_a \over 2 \mu} A(\phi) + {v_0 q} \, \cos \phi -i \left[\Gamma \mu + {1 \over 4} (1+\lambda \cos 2 \phi)^2 \right]{K \over \mu}q^2 \nonumber \\
&=& -{i \over 2 \mu}\left[{\rho v_0^2 \over 2} \cos^2\phi - \sigma_a \cos 2 \phi (1 + \lambda \cos 2 \phi)\right] + O(q, -i q^2)
\end{eqnarray}
and the corresponding twist-bend mode has frequency
\begin{eqnarray}
\label{twist_largeq}
\omega^t &=& -i {\sigma_a \over 2 \mu} A(0)\cos^2 \phi + {v_0 q} \, \cos \phi -i \left[\Gamma \mu + {1 \over 4} (1+\lambda)^2 \cos^2 \phi \right]{K \over \mu}q^2\nonumber \\
&=&  -{i \over 2 \mu}\left[{\rho v_0^2 \over 2}  - (1 + \lambda)\sigma_a\right]\cos^2 \phi + O(q, -i q^2)
\end{eqnarray}
\end{widetext}
\begin{figure}[!h]
	\begin{center}
		\includegraphics[width=0.8\linewidth]{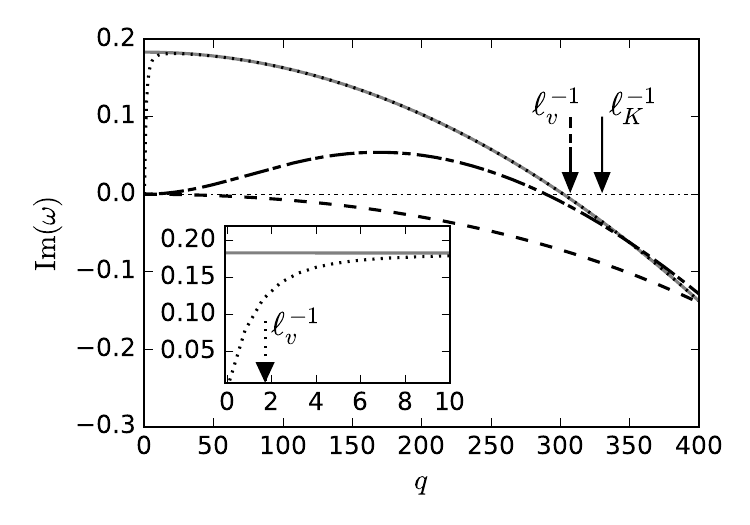}
	\end{center}
	\caption{\label{fig:sim_res1a} Growth rate versus wavenumber. Gray line, Stokesian limit; black dotted line, $O(q)$ unstable, $\ell_v=1$, $R=5 \times 10^{-2}$; black dash-dotted, $O(q^2)$ unstable, $\ell_v=3 \times 10^{-3}$,  $R=4.5 \times 10^{3}$; black dashed line, stable, $\ell_v=3\times 10^{-4}$, $R=4.5 \times 10^{5}$. Arrows indicate  wavenumber corresponding to $\ell_v$ for the unstable cases. For all the dispersion curves we use $\phi=55^{\circ}$ and $K=10^{-6}$ which sets $\ell_K=3.2 \times 10^{-3}$.}
\end{figure}
where $A(\phi)$ is as defined in Eqs.~\eqref{eq:omvsq} and~\eqref{eq:omvsq34}. \ADD{The Stokesian instability of active liquid crystals \cite{SIMSR2002,rmp,Voit1}, with a single growth-rate scale $\sigma_a/\mu$, emerges from Eqs.~\eqref{splay_largeq}, \eqref{twist_largeq} if $\rho$ is set to zero. For $\sigma_a < 0$, contractile or pusher suspensions, which is not the case we are focusing on in this work, Eq.~\eqref{splay_largeq} predicts an instability for $\phi > \pi/4$, which is splay-dominated. For $\sigma_a > 0$, the extensile or pusher case, Eq.~\eqref{splay_largeq} predicts an instability for $\phi < \pi/4$, which is bend-dominated. More important, Eq.~\eqref{twist_largeq} predicts an bend instability for all directions other than pure twist $\phi = \pi/2$, although of course the value of $q$ below which the instability is seen approaches $0$ as $\phi \to \pi/2$.} In general, however, Eqs.~\eqref{splay_largeq} and \eqref{twist_largeq} are not Stokesian expressions but short-wavelength limits of the linearized dynamics of a polar active suspension \textit{with inertia}, which enters through $R$. We see in particular that the stability criteria in this large-$q$ regime are identical to those for the $O(q)$ mode at small $q$. Thus a twist-bend instability, with a growth rate $\sim \sigma_a/\mu$ for $\mbox{max}(\ell_v^{-1},\ell_{\sigma}^{-1}) \ll q \ll \ell_K^{-1}$ takes place if $R < 1+\lambda$. This establishes our claim that the $O(q)$ instability is the small-$q$ extension of the Stokesian instability \cite{SIMSR2002} of active suspensions. The $O(q^2)$ instability that intervenes at small $q$ as $R$ is increased does not reflect itself in the large-$q$ dynamics. Fig. \ref{fig:sim_res1a} displays the growth or decay rates of the twist-bend mode as a function of wavenumber as $R$ is varied~at $\beta=10^{-4}$.

It is important to keep in mind that the active stress $\sigma_a$ is a partial description of the mechanics of self-propulsion based on an estimate of the force-dipole concentration, and is not a priori determined by $v_0$. To take an extreme case, Stokesian swimmers with no force dipole exist, e.g., the pure quadrupole~\cite{Purcell,leshansky,Leoni-Liverpool-film}
Assuming a volume fraction of order unity, let us nonetheless try to estimate $R$ for typical swimmers of speed $v_0$  (the distinction between the speeds of self-propulsion and self-advection being unimportant for this discussion)  and size $b$ (although we must remember that this size is notional in our coarse-grained description). For Reynolds number Re small at the scale of the individual organism it is plausible that $\sigma_a \sim \mu v_0/b$. In that case $R \equiv \rho v_0^2/2 \sigma_a \sim  \rho v_0 b/\mu = \mbox{Re} \ll 1$, so we can replace Eqs.~(\ref{splay_largeq}) and (\ref{twist_largeq}) by their Stokesian approximations. For swimmers at nonzero Reynolds number it is less obvious how to estimate $\sigma_a$. If we take it still to be a viscous stress then $R =\mbox{Re}$ continues to hold, so now $R$ dominates in Eqs.~(\ref{splay_largeq}) and (\ref{twist_largeq}), or in Eqs.~(\ref{eq:omvsq}) and (\ref{eq:omvsq34}), guaranteeing stability. Even if $\sigma_a \sim \rho v_0^2$, $R \sim 1$ and it is plausible that the instability is averted \cite{foot:quad}.

\textit{Dominance of twist in the three-dimensional extensile instability} -- A noteworthy feature, to our knowledge not discussed in the literature, emerges in our three-dimensional analysis: there are two families of bend instability -- mixed with splay as in Eqs.~(\ref{eq:smallq12}) and (\ref{splay_largeq}) and twist as in Eqs.~(\ref{eq:smallq34}) and (\ref{twist_largeq}). Interpolation with bend mitigates the instability in Eqs.~(\ref{eq:smallq12}) and (\ref{splay_largeq}), crossing over to stability for large enough $\phi$, but twist in Eqs.~(\ref{eq:smallq34}) and (\ref{twist_largeq}) has no such effect. The twist-bend instability Eqs.~(\ref{eq:smallq34}) and (\ref{twist_largeq}) should thus dominate, as it occurs for all $\phi$ except precisely $\pi/2$.  This abundance of twisted unstable modes in Eqs.~(\ref{eq:smallq34}) and (\ref{twist_largeq}), independent of the roles of polarity and inertia, is doubtless the explanation of the numerical observations of Shendruk {\it et al.} \cite{shen} in their study of three-dimensional extensile active nematics. 

We summarise this section by noting that, when inertia is taken into account, orientable active suspensions can have two types of linear instability at small wavenumber $q$, governed by the dimensionless parameter $R$ [Eq.~(\ref{Rdef})].
The instability growth-rates are $O(q)$ for $R< R_1 = 1+\lambda$ where $\lambda$ is a flow-alignment parameter and $O(q^2)$ for $R_1< R < R_2 \sim R_1/\beta$ where $\beta$ is defined in Eqs.~(\ref{eq:omvsq}) and~(\ref{eq:omvsq34}). Linearly stable behaviour is found for $R>R_2$. As $\beta \sim 10^{-4}$ in molecular systems, the $O(q^2)$-unstable regime occupies a rather large range in parameter space. Indeed one could argue that the typical behaviour is that corresponding to the $\beta =0$ limit, in which the aligned state is always linearly unstable, either at $O(q)$ or at $O(q^2)$. In \S \ref{sec:numerics} we gain insight beyond this linear analysis through a detailed numerical study to discover the long-time fate of the system in these unstable regimes. 

\section{Numerical studies of active hydrodynamics with inertia} \label{sec:numerics} 
In the following section we describe detailed numerical solutions of our equations, with an emphasis on the changes in behavior as $R$ is varied.  We first verify the predictions of our 
linear stability analysis. Next, for extensile suspensions,  we reveal an inertia-driven nonequilibrium phase transition from a disordered defect-turbulent state for $0<R<R_1$ to an ordered phase-turbulent state for $R_1 \le R < R_2$. We characterize these using the polar order parameter, that is, the macroscopic steady-state average of $\bm{p}$, correlation functions and energy spectra.

\subsection{Direct Numerical Simulations (DNS)}
We numerically integrate Eqs.~\eqref{eq:vel} and~\eqref{eq:dir} in square and cubic domains of volume $\mathcal{L}^d$ in dimensions $d=2$ and $3$. Spatial discretisation, with $N^d$ collocation points, is conducted by employing a
pseudo-spectral method~\cite{canuto}  for Eq.~\eqref{eq:vel} and a fourth-order central finite-difference scheme
for Eq.~\eqref{eq:dir}. For temporal integration we use a second-order Adams-Bashforth scheme \cite{coxmatthews}.
Consistent with the linear stability analysis conducted earlier, we choose a uniform ordered state with transverse 
monochromatic perturbation as the initial condition, i.e. 
$\bm{u}=\bm{0} + A \hat{\bm{e}}_\perp\cos\bm{q} \cdot {\bm r}$, $\bm{p}=\bm{\hat{x}} + B \hat{\bm{e}}_\perp\cos\bm{q} \cdot {\bm r}$ where $\hat{{\bm e}}_\perp \equiv (\hat{\bm y} + \hat{\bm z})/\sqrt{2}$ is a unit vector 
in the plane perpendicular to the ordering direction, and  we have made the arbitrary but acceptable choice $A=B=10^{-3}$. 

We monitor the time-evolution of perturbations and, in the turbulent steady state, investigate the statistical properties of the velocity and the director fields. In Table~\ref{tab}, we summarize the parameters used in our DNS.

\begin{figure}[!h]
    \centering
    \includegraphics[width=0.8\linewidth]{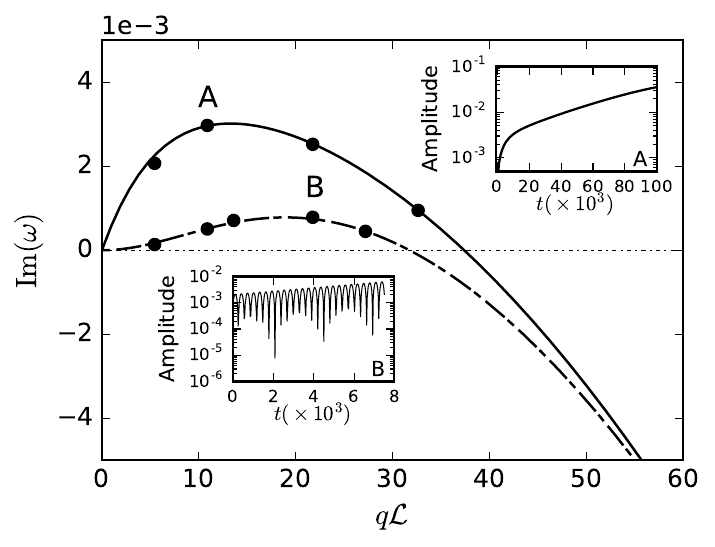}
    \caption{\label{fig:sim_res1}
        Comparison of the growth rates obtained from dispersion relation Eq.~\eqref{eq:omvsq34}  with those from DNS (black
        dots).  (Inset) Initial time-evolution of the perturbation amplitude $|{\bm q}_\perp \times {\delta \bm p}_{\perp q}|$
        for $O(q)$: $R=10^{-2}$ (A), and  $O(q^2)$: $R=4$ (B)  growth rates (run {\tt SPP2}). Note that we choose
        $\phi=55^{\circ}$ for the initial perturbations.
    } 
\end{figure}

\begin{table}[!t]
    \begin{tabular}{@{\extracolsep{\fill}} c  c  c  c  c  c  c }
        \hline
          & D & $\mathcal{L}$ & $N$ & $v_0 (\times 10^{-2})$ & $K (\times 10^{-3})$ & $R \equiv \rho v_0^2/2 \sigma_a$ \\
          \hline \hline
        {\tt SPP1}  & 3   & $2\pi$   & $128$    & $3.16$     & $1$ & $0.02,0.0625$    \\
        {\tt SPP2}  & 3   & $10\pi$  & $160$    & $0.7,13.4$ & $2$ & $0.01,4$         \\
        {\tt SPP3}  & 3   & $10\pi$  & $320$    & $3.16$     & $1$ & $0.1 - 2$        \\
        {\tt SPP4}  & 2   & $20\pi$  & $1024$   & $3.16$     & $1$ & $0.05 - 2.0$     \\
        {\tt SPP5}  & 2   & $32\pi$  & $1024$   & $3.16$    & $1$ & $0.15,0.20$      \\
        {\tt SPP6}  & 2   & $40\pi$  & $2048$   & $3.16$     & $1$ & $0.25,0.30,0.35$ \\
        {\tt SPP7}  & 2   & $64\pi$  & $3072$   & $3.16$     & $1$ & $0.4 - 0.6$      \\
        {\tt SPP8}  & $2$ & $80\pi$  & $4096$ & $3.16$     & $1$ & $0.7$            \\
        {\tt SPP9}  & $2$ & $80\pi$  & $2048$ & $3.16$     & $1$ & $1.25$           \\
        {\tt SPP10} & $2$ & $160\pi$ & $4096$ & $3.16$     & $1$ & $8.0$            \\
        {\tt SPP11} & $2$ & $128\pi$ & $8192$ & $3.16$  & $1$ & $0.01$  \\
        \hline
    \end{tabular}
    \caption{\label{tab}
        Spatial dimension $D$ of the domain and parameters $\mathcal{L},~N,~v_0,~K$, and $R$ used in our direct numerical simulations. The suspension density $\rho=1$, $\lambda=0.1$, $\mu=0.1$ and the rotational mobility $\Gamma=1$ are kept fixed for all the runs. \ADD{Note that as $R$ approaches $R_2$ the range of linearly unstable modes shrinks and is restricted to small wave-numbers, i.e., large length-scales. To resolve these unstable modes as well as the small-scale fluctuations that arise because of the nonlinear couplings, for $R=8$ we use a square domain with each side of length $160\pi$ and discretize it with $4096^2$ collocation points.}}
\end{table}

\subsection{Initial growth of instabilities}
We now present a comparison between the short-time growth obtained from the DNS with the analytical predictions of the linear stability analysis.
The plot of the bend-twist dispersion curve given by Eq.~\eqref{eq:omvsq34} for $\phi=55^{\circ}$ is shown in Fig.~\ref{fig:sim_res1}.  The black dots indicate the the initial temporal growth rate of perturbations obtained from our DNS, which shows excellent agreement with the analytical results. Furthermore, our simulations correctly capture the exponential and oscillatory characters of the growth for $R<R_1$ and $R_1<R<R_2$ respectively.  Note that for $R<R_1$, the exponential growth rate of perturbations is much faster than the oscillatory  kinematic contribution ${\rm Re}[\omega]= v_0 q \cos\phi$.  For $R_1<R<R_2$, $\rm{Re}[\omega]$  has contributions from both the kinematic and the inertial terms. Therefore, we observe an exponential growth of $|\bm{q_\perp}\times \bm{\delta p}_{\perp {\bm q}}|$ for $R<R_1$ [see \subfig{fig:sim_res1}{A}], but oscillatory growth for $R_1<R<R_2$ [see \subfig{fig:sim_res1}{B}].

\subsection{A flocking phase transition}

\begin{figure*}[!t]
\includegraphics[width=0.99\linewidth]{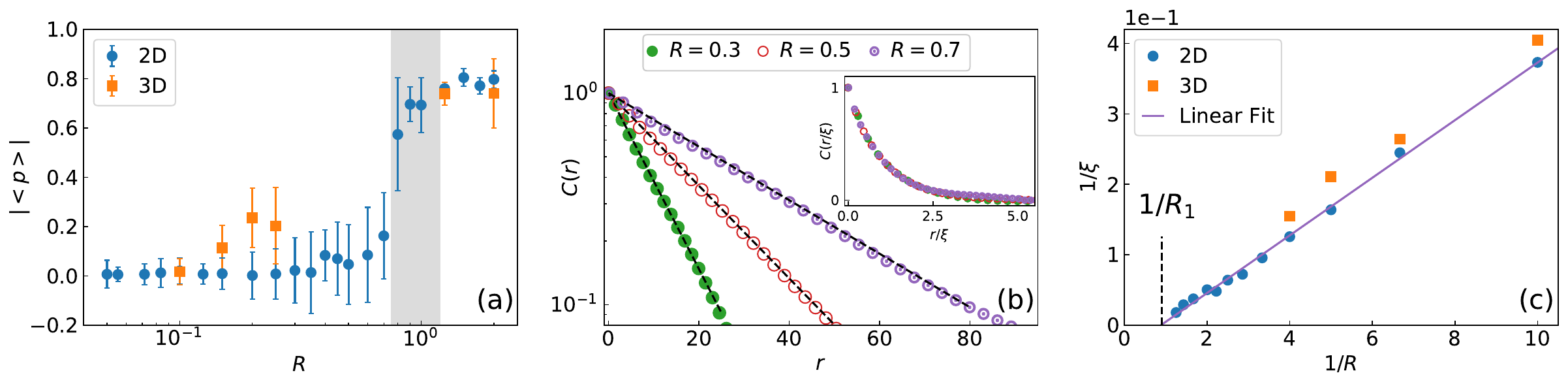}
\caption{\label{fig:opvsri}
    (a) Variation of the order parameter $|\langle {\bm p} \rangle|$ with $R$ for our 2D and 3D simulations (see
    Table~\ref{tab}), with the shaded region indicating the transition regime around $R=R_1$ as predicted by the linear
    stability analysis. For each data point, the spatio-temporal  average is calculated from about $60$ statistically
    independent realizations and the standard deviation about the average is shown as the error-bar. (b) Semilog plot of
    the correlation function $C(r)$ versus $r$ for $R=0.3,0.5,$ and $0.7$ ($R < R_1$, runs ${\tt SPP6-8}$). Dashed black
    lines indicate the exponential fit. Inset: Collapse of steady state correlation function when distance is scaled
    with the correlation length. (c) Plot of inverse correlation length $1/\xi$ versus $1/R$. Continuous purple line
    shows the linear fit to 2D data. Note that from the intercept of the linear fit on the horizontal axis we conclude
    that the correlation length diverges around $R = R_1$.
}
\end{figure*}

\begin{figure*}[!t]
\begin{center}
\includegraphics[width=\linewidth]{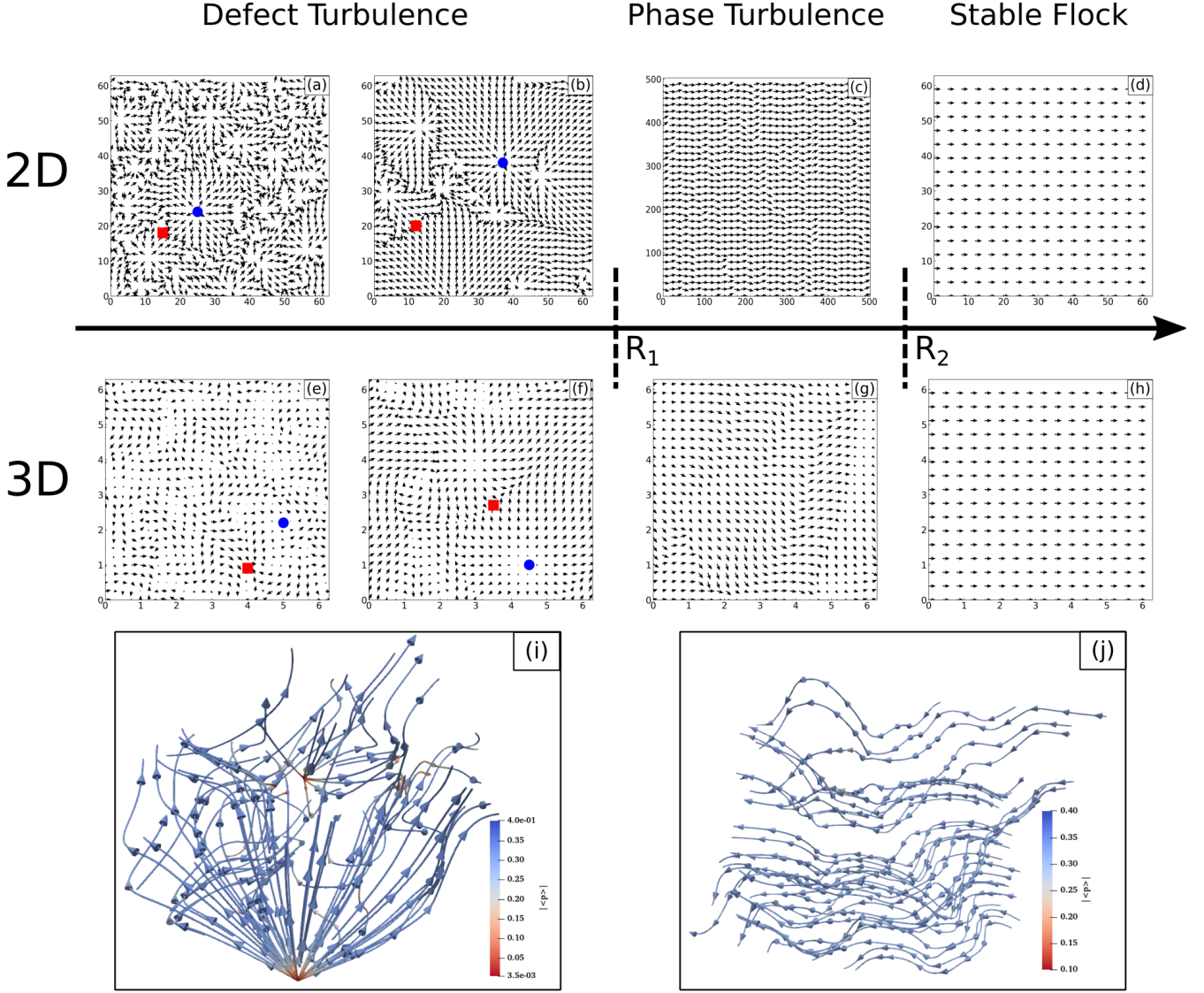}
\caption{\label{fig:phasedgm} Increasing inter-defect distance as function of $R$. Order parameter streamlines 
for $2$D (runs {\tt SPP4}, {\tt SPP6} and {\tt SPP9}): (a) $R=0.1$, (b) $R=0.25$, (c) $R=8$, and (d) $R>12$;  streamlines in $y=0$ plane for $3$D (runs {\tt SPP1} and {\tt SPP3}): (e) $R=0.02$, (f) $R=0.0625$, (g) $R=1.25$, and (h) $R>12$. Typical hedgehogs are marked with filled black circles and 
red squares indicate saddles. (i) Zoomed-in view of the three-dimensional order parameter streamlines showing the complex patterns between a hedgehog-saddle-hedgehog configuration in (f). (j) Three-dimensional nearly ordered configuration in the phase-turbulent regime in (g).}
\end{center}
\end{figure*}
We now investigate the morphology and statistical properties of the orientation and flow emerging from the instabilities discussed above. \Fig{fig:phasedgm} shows the typical flow structures observed in our DNS with increasing $R$ in the statistically steady state. For $0<R<R_1$, we observe hedgehog defects. 
The inter-defect spacing grows with increasing $R$. Unexpectedly, when $R$ increases past the first threshold $R_1$, a fluctuating but on average aligned state emerges. As we remarked in the Introduction, this is clear numerical evidence that $R=R_1$ marks a nonequilibrium phase transition from a statistically isotropic state to a flock or, in the terminology of spatiotemporal chaos, from defect turbulence to phase turbulence~\cite{egolf,Hayot,chatemanneville}. In the latter state long-wavelength statistical variation of the broken-symmetry variable is present but the amplitude of the order parameter is not destroyed by defects. We have not, however, measured the system-size dependence of the positive Lyapounov spectrum to establish spatiotemporal chaos quantitatively. \ADD{We do not know the mechanism that serves to preserve macroscopic flocking order despite the $O(q^2)$ instability. It appears that the growing amplitude of perturbations at small wavenumber $q$ triggers nonlinear effects which couples to large $q$ where the dynamics is stable. The behavior is reminiscent of that reported by Jayaprakash et al. \cite{Hayot} for the Kuramoto-Sivashinsky (KS) equation. The KS equation is a deterministic partial differential equation (PDE) with a negative diffusivity and hence a linear instability with growth rate $\propto q^2$ at small wavenumber $q$, peaking at a wavenumber $q_*$, stable behavior at large $q$ thanks to terms at higher order in $q$, and a nonlinearity that transfers weight from small to large $q$. Hayot et al. \cite{Hayot} carry out a numerical coarse-graining, i.e., a spatial low-pass filtering, on the two-dimensional KS equation to show that the effective equations of motion for the modes with $q<q_*$ are those of a \textit{stochastic} PDE with a \textit{positive} diffusivity. It is possible that such a mechanism is at work in our case, but to settle this issue will require a treatment analogous to that of \cite{Hayot} for our substantially more complicated equations.
}. 

We now focus on the properties of the nonequilibrium phase transition. In Fig.~\ref{fig:opvsri}(a), we plot the magnitude $|\langle {\bm p}\rangle|$ of the polar order parameter in the statistically steady state with increasing $R$, where angle brackets $\langle \cdot \rangle$ denote spatio-temporal averaging. For $R<R_1$,  $|\langle {\bm p} \rangle|$ is consistent with zero. We observe an onset of polar order once $R$ increases beyond $R_1\equiv 1+\lambda$. Fig.~\ref{fig:opvsri}(a) shows the order parameter for the largest system sizes studied; at large $R$ the values at half that system size are very similar. However, a detailed finite-size scaling analysis needs to be undertaken to find the correct scaling near the critical region \cite{FSS}. 

In the defect-turbulence regime, we study the steady-state correlation function $C(r)=\langle {\bm p}({\bm x} +
    \bm{r}) \cdot {\bm p}({\bm x})\rangle/\langle {{\bm p}(0)}^2 \rangle$, where the angular brackets denote spatial averaging. We plot the correlation function $C(r)$ versus $r$ in
    Fig.~\ref{fig:opvsri}(b) and evaluate the correlation length by fitting an exponential decay $\exp(-r/\xi)$ to the numerical data \footnote{The value of $\xi$ obtained from the fit is comparable to the one obtained using the definition $\xi \equiv \int_0^{\mathcal{L}/2} C(r) dr$.}. We see that the correlation functions for different values of $R<R_1$ fall on a single curve plotted against $r/\xi$. Moreover, from Fig. \ref{fig:opvsri}(c), $\xi$ grows and possibly diverges as $R \to R_1$; our limited data points are consistent with an exponent of unity. Further progress requires finite-size scaling studies and measurements of order-parameter correlations at asymptotically small wavenumber \cite{FSS} for $R>R_1$ to test the nature of the ordered state. 

\subsection{Energy spectrum}
\begin{figure*}[!t]
\includegraphics[width=\linewidth]{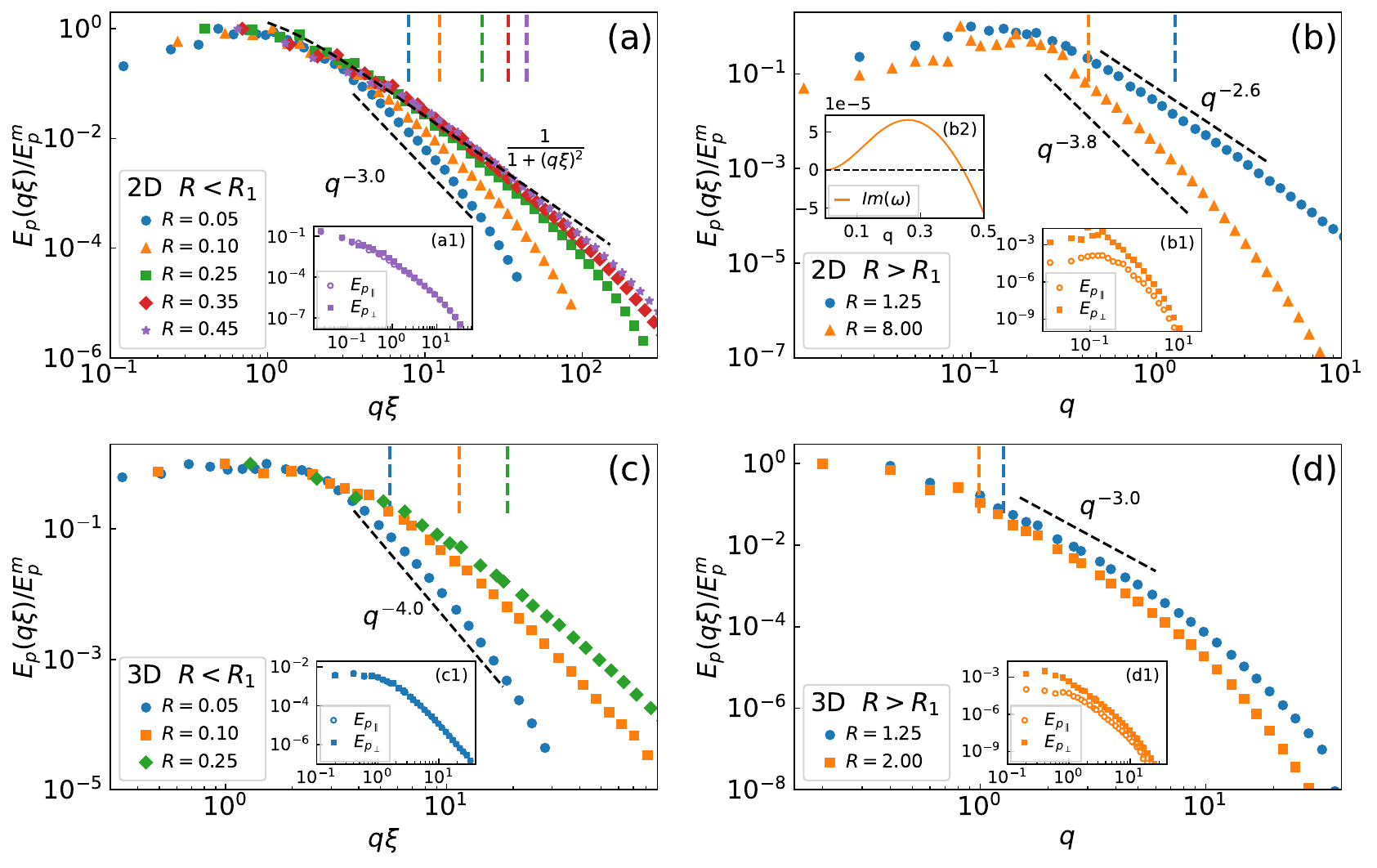}
\caption{\label{fig:turb1} Order parameter energy spectrum $E_p(q)$ for different values of $R$ for (a,b) two-dimensional [runs {\tt SPP4}, {\tt SPP6}, {\tt SPP9} and {\tt SPP10}] 
and  (c,d) three-dimensional [runs {\tt SPP3}] active suspension. \ADD{For $R<R_1$ and $q\xi>1$, we observe  a Porod's tail due to defects, i.e.  $E_p(q)\sim q^{-3}$ in 2D and $E_p(q)\sim q^{-4}$ in 3D. As $R$ approaches $R_1$, we find $E_p(q)\sim 1/[1+ (q \xi)^2]$ consistent with the exponential decay of the correlation function.}  For $R_1<R< R_2$, $E_p(q)\sim q^{-2.6}$ for $R=1.25$ and the slope increases to $q^{-3.8}$ for $R=8$. For different values of $R$, dashed vertical lines (with same color as markers) indicate the largest $q$ which is linearly unstable. Insets (a1), (b1), (c1), and (d1) show the spectra $E_{p_\parallel}$ for components of $\bm{p}$ along the mean ordering direction and $E_{p_{\perp}}$ for a representative direction orthogonal to it. For $R<R_1$, fluctuations are isotropic whereas for $R>R_1$, transverse fluctuations dominate, with $E_{p_{\perp}} \approx 10^2 E_{p_{\parallel}}$. Note that although the mean order parameter is of course consistent with zero for $R<R_1$, we have used the numerically measured mean ordering direction to define $||$ and $\perp$.  Inset (b2) shows the growth rate for $R=8$. Note that only a small number of modes between $q=0$ and $q=0.5$ are linearly unstable.}
\end{figure*}
\begin{figure*}[!t]
\includegraphics[width=\linewidth]{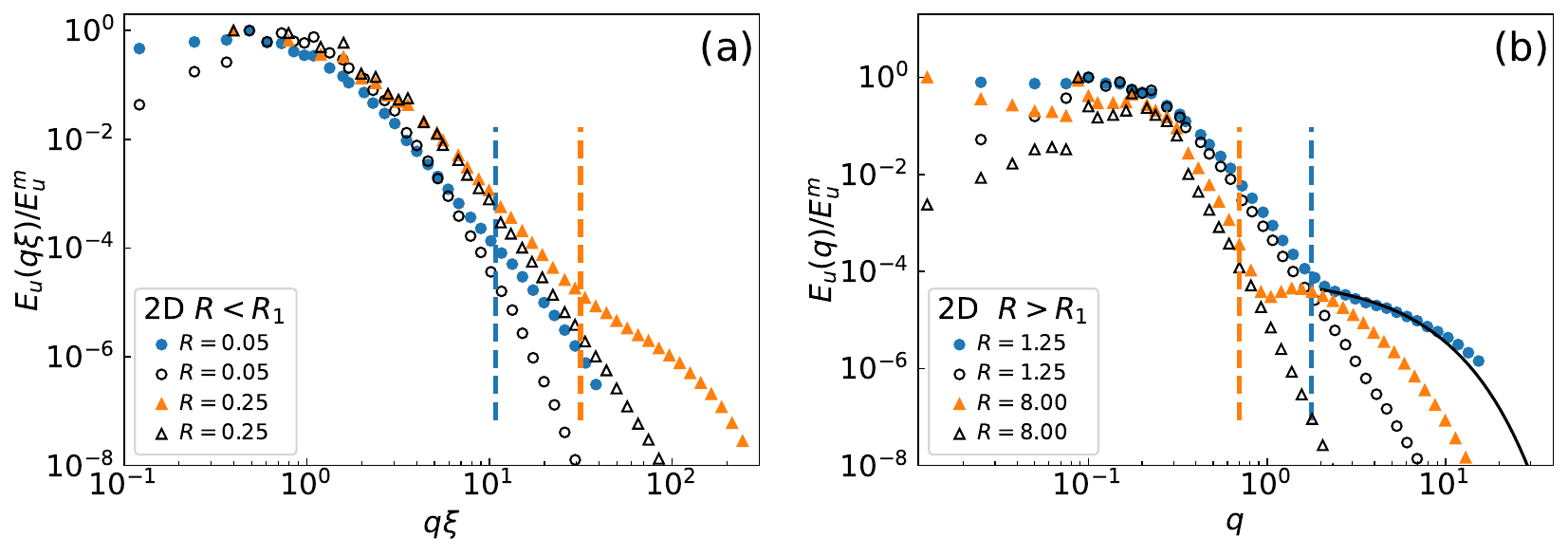}
\caption{\label{fig:turb2} Kinetic energy spectrum $E_u(q)$ (filled symbols) for different values of $R$ for 2D active suspension [runs {\tt SPP4}, {\tt SPP9} and {\tt SPP10}] for (a) $R<R_1$ and (b) $R>R_1$ . Similar to order parameter spectrum we observe power-law behaviour for $1<q\xi< q_\sigma \xi$, where $q_\sigma\equiv 2\pi/\ell_\sigma$. For small-$q$, we find a good agreement between the energy spectrum and the prediction $E_u(q)\sim\left[E_p(q)\right]^2$  (unfilled symbols). For $R>R_1$ and large-$q$ ($q>q_\sigma$) the energy spectrum shows an  exponential decay $E_u(q) \sim \exp(-0.31 q)$ (black line). For different values of $R$, dashed vertical lines (with same color as markers) are drawn at  $q=q_\sigma$.}
\end{figure*}

\ADD{A state of complex, correlated but disorderly flow is seen in a wide variety of suspensions of motile organisms and motorized biofilaments. It has been termed active turbulence and analyzed through the study of energy spectra as in conventional turbulence \cite{alert2021active,giomi2,wensink1,Ishikawa1,wol2008,shelley2,giomi3,URZAY2017,sanchez,Li2019,Joanny2,bratanov}. However, these studies have all considered systems with negligible inertia. Here we examine numerically the spatial power spectral densities for the polar order parameter and the hydrodynamic velocity field, in the defect- and phase-turbulent regimes -- the latter owing its existence to inertia. In keeping with typical turbulence studies, we use the shell-averaged energy spectra of the velocity and the order parameter
\begin{eqnarray} \label{eq:shellspectra} 
\nonumber
E_{u}(q) &=& \sum_{q-1/2 \leq |{\bm m}| < q+1/2} |{\bm u}_{\bm m}|^2,~{\rm and}~\\
E_{p}(q) &=& \sum_{q-1/2 \leq |{\bm m}| < q+1/2} |{\bm p}_{\bm m}|^2,
\end{eqnarray}
where ${\bm u}_{\bm m}$ and ${\bm p}_{\bm m}$ are the Fourier coefficients of the velocity ${\bm u}$ and order parameter ${\bm p}$ fields.  \\
 Among the features of interest are Porod's Law regimes corresponding to the fields of topological defects. In addition, for $R<R_1$ we find velocity correlations of Ornstein-Zernike form, with a correlation length much larger than that of the order parameter, whose origin we discuss below. For $R>R_1$, the phase-turbulent but ordered state, we present preliminary evidence of fluctuations of the broken-symmetry or Nambu-Goldstone mode.} 
The behaviours of $E_p(q)$ and $E_u(q)$ for a range of values of $R$ are displayed in Figs.~\ref{fig:turb1} and \ref{fig:turb2}.

\ADD{{\it Energy spectra of the order parameter}}--  We observe that for $R<R_1$, the spectrum peaks around $q\xi\sim 1$. \ADD{At moderate values of $R$, $0.1 < R<1$, because of exponential decay in orientational correlations, we expect $E_p(q)\sim1/(1+ \xi^2 q^2)$ for $q \xi>1$. On the other hand, for $R \ll 1$, because of defects, we expect a Porod's scaling~\cite{bray2002} $E_p(q)=q^{-3}$ in two dimensions and $E_p(q)=q^{-4}$ in three dimensions for $q\xi \gg 1$. Recent studies on dry active matter \cite{rana2020} using a scale-by-scale budget analysis revealed that, even in the presence of defects, the nonlinear transfer mechanisms could lead to a non-Porod scaling. Unfortunately, we do not have sufficient scaling range -- due to  modest grid-resolution at $R\ll R_1$ -- to undertake such analysis.} We emphasize that the quoted exponent values are empirically determined by conservatively selecting an appropriate dynamic range of wavenumbers away from the smallest ($\sim 1/L$) and the largest, viz., $q_K\equiv 2\pi/\ell_K$, beyond which Frank  elasticity dominates.

\ADD{In the phase-turbulent regime, $R_1<R<R_2$, we observe $E_p(q) \sim q^{-3}$ for $R$ close to $R_1$. As $R$ approaches $R_2$,  the range of linearly unstable modes shrinks and is restricted to wave-numbers close to large scales [small $q$, see Fig. \ref{fig:turb1}(b)].  For the linearly unstable modes we observe a weak $q$-dependence, whereas for wave-numbers outside the linearly unstable regime, the nonlinearities lead to a transfer of order-parameter fluctuations to small scales with a power-law spectrum $E_p(q) \sim q^{-3.8}$ [see Figs.~\ref{fig:turb1}(b) and (d)].}

\ADD{{\it Energy spectra of the velocity}} -- In \Eq{eq:vel} we expect the dominant balance to be between acceleration and activity as the Reynolds number obtained by comparing the advective and viscous terms, based on the root-mean-square hydrodynamic velocity, is small ($\Rey \equiv \rho u_{rms} \xi/\mu \le 0.5$) \footnote{We have verified that the results of our DNS do not change if the advective nonlinear term in \Eq{eq:vel} is absent.}. We therefore expect for small $q$, $\omega u_q \sim \sigma_a q \sum_k {\bm p}_k {\bm p}_{q-k}$. If we assume that the dominant contribution to the convolution comes from terms with $|{\bm k}| = |{\bm q} - {\bm k}| =q$, i.e., on the same shell in Fourier space, we get $\omega u_q \sim \sigma_a q E_p(q)$. Using $\omega \sim v_0 q$ [see \Eq{eq:omvsq}] we get, $E_u(q) \sim (\sigma_a/v_0)^2 E_p(q)^2$.  The plot in Fig.~\ref{fig:turb2}(a) shows good agreement between $E_u(q)$ obtained from our DNS and the conjecture above for small $q$. For large $q>2\pi/\ell_\sigma$, we expect viscous dissipation to be dominant and therefore, similar to the  dissipation range in hydrodynamic turbulence, we expect an exponential decay in the energy spectrum $E_u(q) \sim \exp(-a k^\delta)$ \cite{smi90,Fri96}. From our numerical simulations, we find $\delta=1$.

It is worth noting that although turbulence in an apolar active suspension is controlled by half-integer defects in 2D \cite{giomi3,URZAY2017} and disclination loops or line defects in 3D \cite{ducl20,rav20}, the flow energy spectrum $E_u(q) \sim q^{-3.5}-q^{-5}$ reported in those works on active nematics does not differ drastically from our observation $E_u(q) \sim q^{-4.8}-q^{-6}$ for defect turbulence in the $R<R_1$ regime in our polar system.

\ADD{\subsubsection{Energy spectra for $R\ll R_1$}
For $R \ll R_1$, the inter-defect separation $\xi$ is comparable to $\ell_\sigma$; however, for length scales much larger than $\xi$, the system should in effect be an unsteady Stokes fluid with fluctuating stresses with short-ranged spatial correlation, and with a correlation time $\tau$. In such a scenario it is straightforward to show \cite{lau2009fluctuating,IPMpc} that the equal-time velocity correlator has Ornstein-Zernike form, so that $E_{u}(q)\sim q^{D-1}/[(q \ell_\tau)^2 +1]$ for $q\xi\ll1$, where $\ell_\tau=\sqrt{\mu \tau/\rho}$ is the distance vorticity diffuses in a time $\tau$. Note: (a) although the analysis of ref. \cite{lau2009fluctuating} contains this result, they do not emphasize the distinct roles of $\xi$ and $\ell_\tau$. (b) The power-law correlations discussed in the inertia-less treatment of  \cite{underhill2008diffusion,Joanny2} amount to the $q \ell_\tau \ll 1$ regime of the above.}

\begin{figure}
    \centering
    \includegraphics[width=\linewidth]{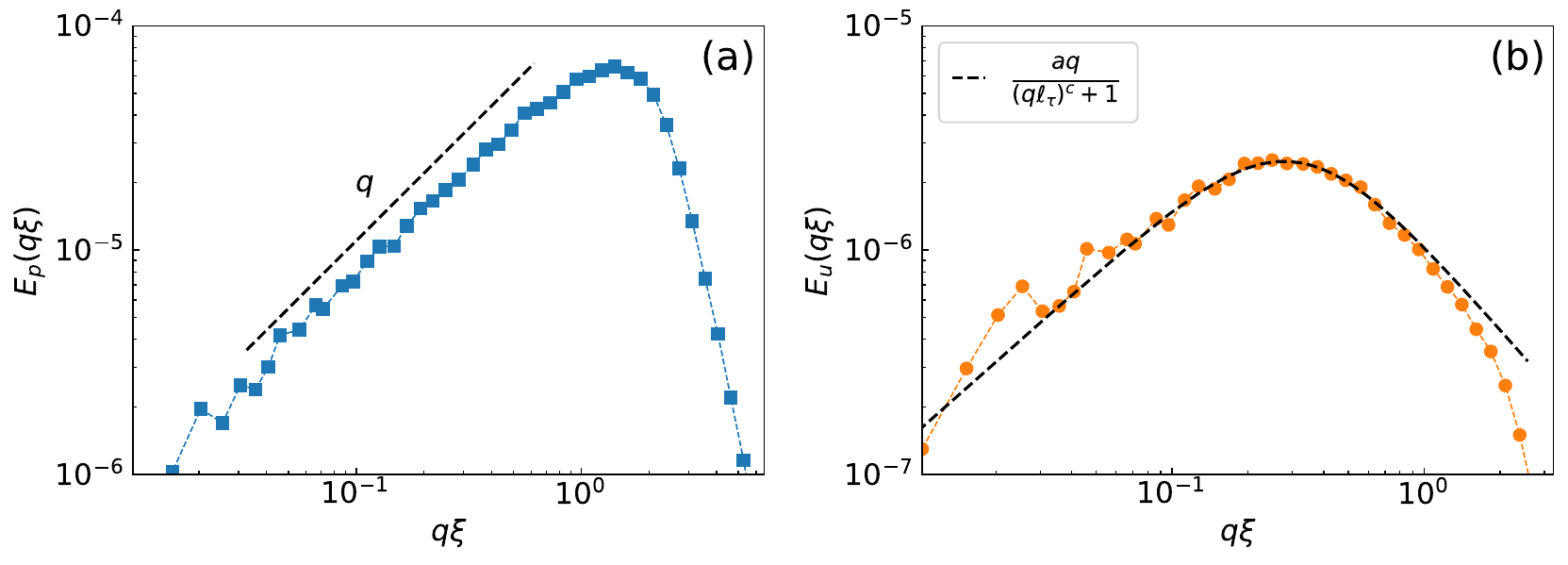}
    \caption{\label{spec_R0p01} \ADD{Energy spectra for $R=0.01$. (a) Order parameter energy spectra $E_p(q\xi)$. The black dashed line indicates $E_p(q)\sim q$ scaling. (b) Kinetic energy spectra $E_u(q \xi)$. A least squares fit to the curve $aq/[(q \ell_\tau)^c +1]$ gives $a=1.6\times10^{-5}$, $c=2.3$, and $\ell_\tau=1$. }}
\end{figure}

\ADD{To investigate this regime, we perform high resolution simulations in two dimensions with large system size and small $R=0.01$ (run {\tt SPP11}) to ensure ${\mathcal L} \gg \xi$.  The plot in Fig.~\ref{spec_R0p01}(a) shows that the $E_{p}(q) \sim q$ for $q\xi \ll 1$ indicating that order parameter fluctuations are  uncorrelated. We, therefore, expect the active stresses to be  spatially uncorrelated. Consistent with the arguments above, the plot in Fig.~\ref{spec_R0p01}(b) shows that the kinetic energy spectrum follows $E_u(q) \sim q/[(q \ell_\tau)^c +1]$ with $c \approx 2.3$ (obtained from a least-squares fit) close to  the theoretically predicted value $c=2$. Note that the length scale $\ell_\tau$ is larger than the interdefect separation $\ell_\tau/\xi\approx3$.  As $R\to 0 $, we expect $\xi\to0$ and $\ell_\tau \gg \xi$; therefore, the peak of spectra in Fig.~\ref{spec_R0p01}(b) would  shift to very small $q\sim1/\ell_\tau$. Thus our analysis naturally recovers and explains the recently observed $E_u(q)\sim q^{-1}$ scaling of active nematic turbulence in the Stokesian regime \cite{Joanny2}.}

\ADD{\subsubsection{Evidence for the broken-symmetry mode?} \label{subsub:BS} We close our discussion of energy spectra with a speculation backed by qualitative numerical measurements. In any noisy ordered state in which a continuous symmetry has been spontaneously broken, the spatial power spectral density should contain information about the broken-symmetry modes, i.e., the components of the order parameter field perpendicular to the mean ordering direction, whose variance should diverge at small wavenumber \cite{Mazenko_noneq_book,MPP,forster1975hydrodynamic}. This variance should be seen in the energy spectrum in the ordered but noisy phase-turbulent state we observe for $R_1<R<R_2$. We offer preliminary evidence for such fluctuations. Insets b1 and d1 to Fig. \ref{fig:turb1} show that for $R>R_1$ the contributions to the energy spectrum from components of $\bm{p}$ in a representative ($\perp$) direction perpendicular to the mean direction of ordering far outweigh those from components in the ordering direction, especially at small $q$. This is consistent with the expectations of an enhanced variance mentioned above. We check for consistency that the spectra for the disordered phase for $R<R_1$ (insets a1 and c1 to Fig. \ref{fig:turb1}) show no such anisotropy.} 

\ADD{A quantitative study of the spectrum of fluctuations at small wavenumber, to test whether the regime $R_1<R<R_2$ has the classic features of a broken-symmetry phase, makes high computational demands. The wavelength at which the linear instability growth rate is maximum can be viewed as the scale of energy injection, and therefore as the small-scale cutoff for a long-wavelength study of the ordered phase. At the same time, scales substantially smaller than this cutoff must be resolved so that the instability and the nonlinear effects leading to phase turbulence can operate. If such a simulation is realized, the approach of choice would be to emulate \cite{Hayot} and construct an effective stochastically-forced theory for the small-$q$ modes via numerical coarse-graining.}

\section{Summary and prospect}
\label{sec:concl}
We have shown that extensile active polar liquid crystals and swimmer suspensions can outswim their viscous instability. Their fate is governed by a control parameter $R$, the ratio of the inertia of self-advection to the scale of active stress. Our stability analysis and numerical studies find evidence for a continuous flocking transition with a growing correlation length as $R$ increases past a threshold of order unity, from hedgehog-defect turbulence to a noisy but \textit{ordered} phase-turbulent state. A quiescent, linearly stable ordered state sets in at larger $R$. These dramatic advances in the theory of flocks in fluid, whose instability \cite{SIMSR2002} can now be seen as simply the Stokesian limit of a rich phase diagram, should stimulate a new wave of experiments on swimmers at nonzero Reynolds number. Important directions for the near future are studies of finite-size scaling and long-wavelength order-parameter correlations for $R>R_1$ to establish the nature of the ordered phase; the construction of an effective stochastic theory for the long-wavelength modes, as carried out \cite{Hayot} for the Kuramoto-Sivashinsky \cite{kuramotosuzuki,sivashinsky} equation; the inclusion of active-particle concentration; the contribution of other polar terms to the dynamics; and the effect of added random forcing on our phase diagram. Meanwhile, we look forward to tests of our theory in experiments on collections of swimmers at small but nonzero Reynolds number \cite{klotsa_persp}, as well as particle-based numerical simulations featuring, for example, collections of ``spherobots'' \cite{dap01,dap02}.

\begin{acknowledgements}
    S.R. was supported by a J. C. Bose Fellowship of the SERB (India) and by the Tata Education and Development Trust.
    R.C. acknowledges the support of the SERB (India). R.C., N.R., and P.P. acknowledge  support from the Department of Atomic Energy (DAE), Government of India, under Project Identification No. RTI 4007. The simulations were performed using resources provided by TIFR, Hyderabad. We thank H Chat\'{e} for valuable comments on an early draft. R.C., N.R., P.P., and S.R. acknowledge the contributions of Aditi Simha, now sadly deceased, to the development and completion of this work.
\end{acknowledgements}

\appendix
\label{app}
\section{Stability analysis with concentration}
 We now present the linear stability analysis in presence of the concentration field. Linearizing  Eqs.~\eqref{eq:vel},~\eqref{eq:dir} and~\eqref{eq:conc}  about the base state $(\bm{u}=0,\bm{p}=\hat{\bm x},c=1)$ we get:
\begin{align}
  \left(\rho\partial_t + \mu q^2\right)\delta \uu_{\perp \qq}
  &= -i\Tq \cdot  \left[\left(\sigma_a +{\lambda-1 \over 2} K q^2\right)\hat{\bm x} \qq_{\perp}
     +q_x \left(\sigma_a + {\lambda + 1 \over 2}K  q^2\right)\bsf{I}\right] \cdot\delta \bm{p_{\perp \qq}}
     -i \Tq \cdot \left(\sigma_a  q_x  \hat{\bm x}\right) \delta c_{\bm q},\label{eq:lu_wc}\\	
 \partial_t \delta \bm{p}_{\perp \qq}
  &= +i\left({\lambda+1 \over 2}q_x \bsf{I}
     -{\lambda-1\over 2} {\qq_{\perp}\qq_{\perp} \over q_x}\right)\cdot\delta \bm{u}_{\perp \qq}
     -\left(iv_0q_x + \Gamma K q^2\right)\delta \bm{p}_{\perp \qq}
     +i\qq_{\perp} E\delta c_{\qq}, \label{eq:lp_wc}\\   
  (\partial_t + iv_1q_x)\delta c_{\bm q}
  &= -i v_1 \bm{q_{\perp}\cdot\delta p_{\perp q}}. \label{eq:lp_cc}
\end{align} 

\begin{figure*}[!h]
    \includegraphics[width=0.49\linewidth]{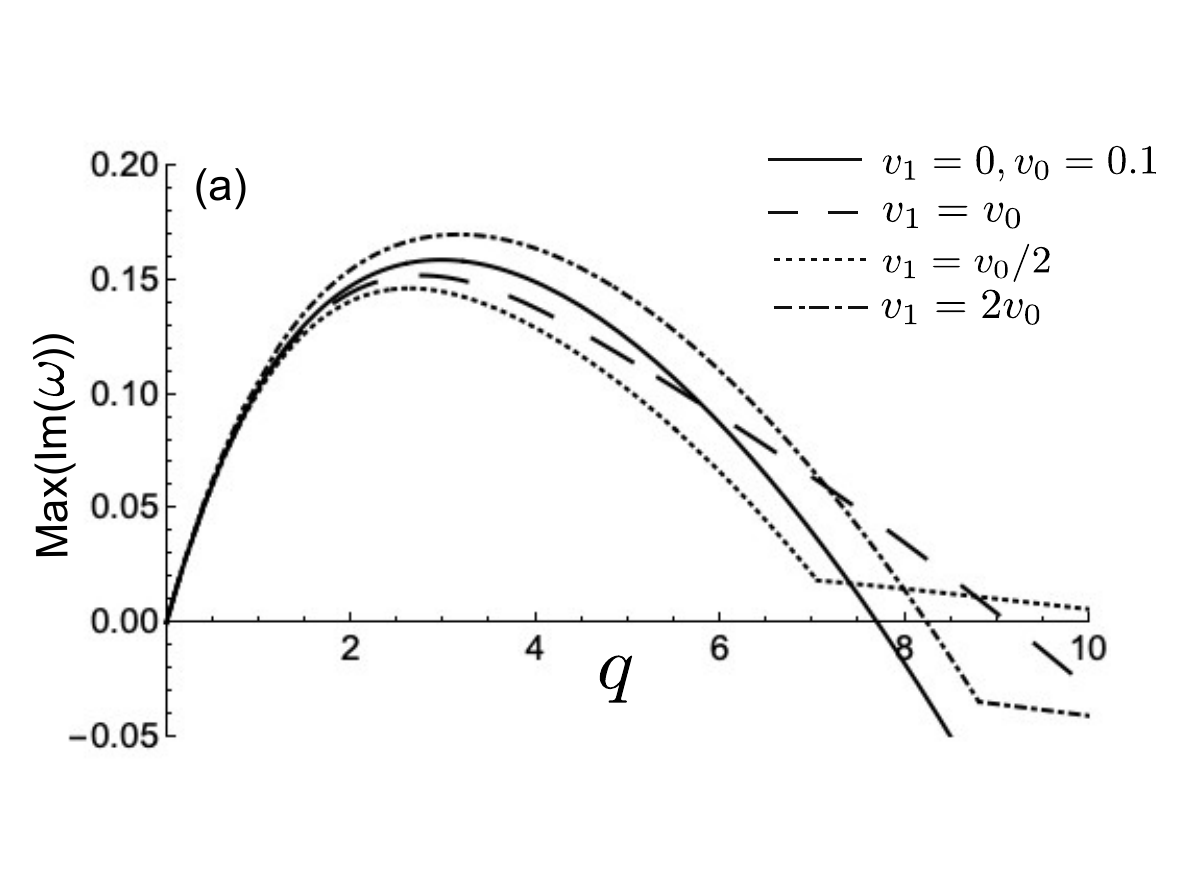}
    \includegraphics[width=0.49\linewidth]{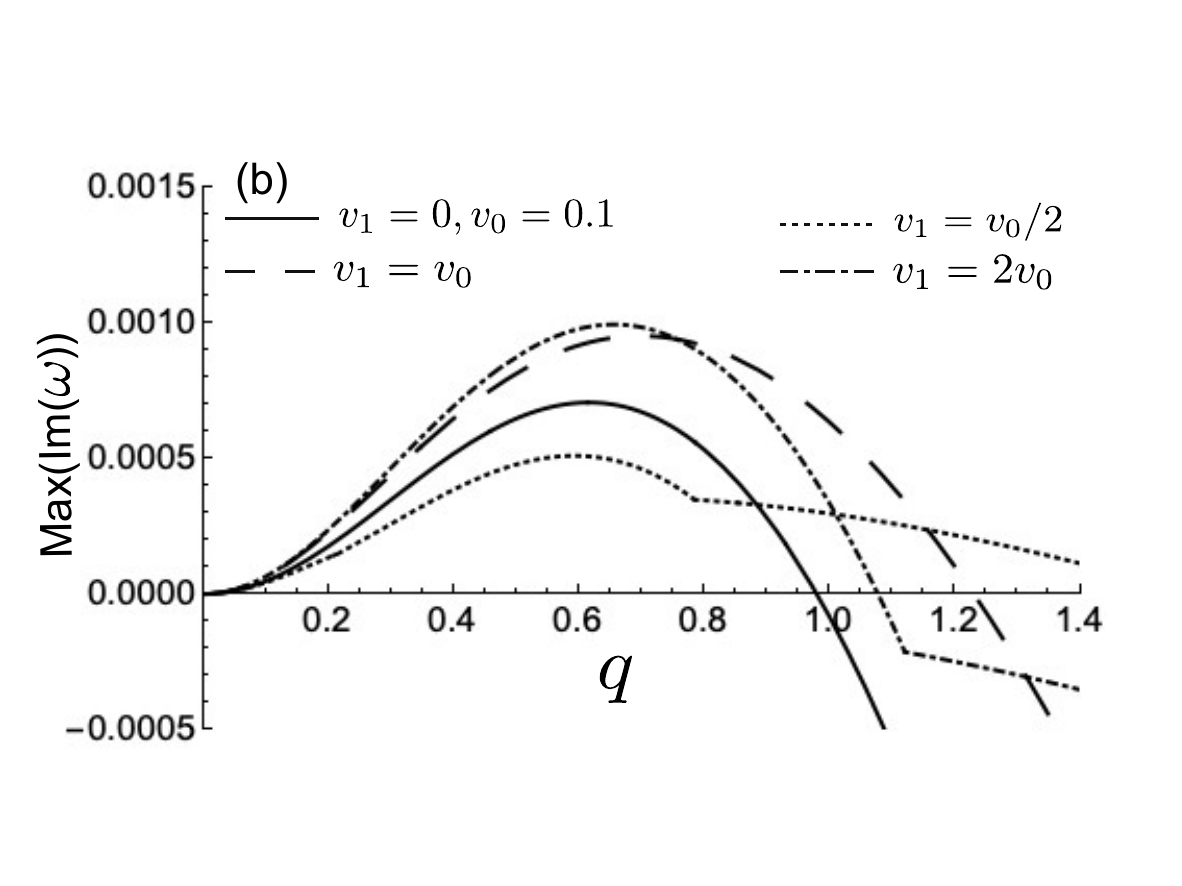}
    \caption{\label{fig:concsplay}
        Maximum growth rate of the unstable splay-bend modes for different values of $v_0,v_1$ for (a) $O(q):$
        $R=0.1$, and (b) $O(q^2):$ $R=5$ regimes. Note that the dispersion relation for $v_1=0$ is identical to  Eq.~\eqref{eq:omvsq} (see discussion below). The other parameters are: $\phi=15^{\circ}$, $\mu=10^{-1}$, $\Gamma
        K=10^{-3}$, $\lambda=0.1$, $E=0.05$.}
\end{figure*}

The dispersion relation for \cref{eq:lu_wc,eq:lp_wc,eq:lp_cc}  are obtained using the same procedure as highlighted in Section~\ref{subsec:linstab}.  The growth rate of the twist-bend modes is same as  Eq.~\eqref{eq:omvsq34} because the terms containing concentration fluctuations in \cref{eq:lu_wc,eq:lp_wc} point in the direction of ${\bm q}_\perp$. The growth rate of the splay-bend modes is identical to Eq.~\eqref{eq:omvsq} for $v_1=0$ because concentration fluctuations decouple from the orientation and velocity distortions. For $v_1\ne 0$ the splay-bend modes couple to the concentration fluctuations and are obtained by taking in-plane divergence ($\nabla_{\perp}\cdot$) on \cref{eq:lu_wc,eq:lp_wc}. We compare the most unstable growth rate  in Fig.~\ref{fig:concsplay} and  show that the $O(q)$ and the $O(q^2)$ behavior at remains unaltered small-q.
\section{Stability analysis including the neglected polar terms}
\begin{figure*}[!h]
	\includegraphics[width=0.55\linewidth]{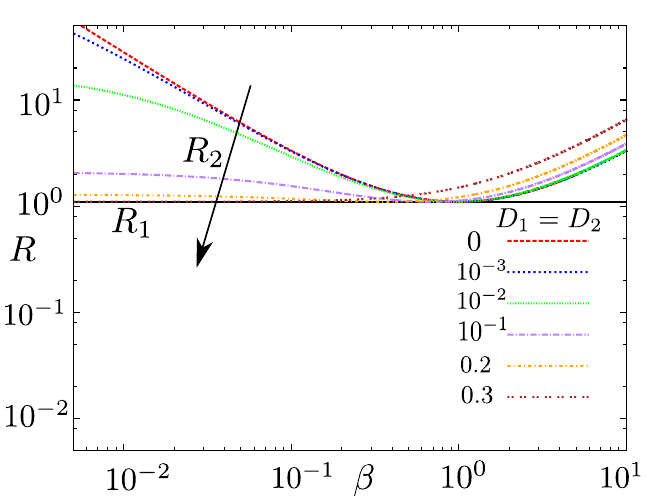}
\caption{\label{fig:Rvsbetapolar} $R$-$\beta$ stability diagram including polar contributions $\ell$ and $\gamma_a$ to flow alignment and active stress.The lines at $R=R_1$ denoting the junction between the $O(q)$ and the $O(q^2)$ instabilities remains unchanged. However the line $R=R_2$ which marks the borderline between the $O(q^2)$ instability and the stable aligned state starts to deviate as the active capillary numbers decrease from very large values towards values of order unity.}
\end{figure*}
We now explain the role of the additional polar terms on the stability thresholds. We first recall the complete
Eqs.~\eqref{eq:vel} and~\eqref{eq:dir}  with the polar terms:
\begin{eqnarray*} 
\rho (\partial_t  \uu +\uu \cdot \nabla \uu) &=& -\nabla P + \mu \nabla^2 \uu + \nabla \cdot ({\bm \Sigma}^a+{\bm \Sigma}^r),~\text{and}\\
\partial_t\bm{p} + (\uu + v_0{\bm p})\cdot\nabla {\bm p}&=& \lambda {\bm S}\cdot\bm{p}+{\bm \Omega}\cdot{\bm p} + \Gamma {\bm h} +\ell \nabla^2 {\uu}, 
\end{eqnarray*}
where ${\bm \Sigma}^a \equiv -\sigma_a \bm{p}\bm{p} - \gamma_a (\nabla \bm{p} + \nabla \bm{p}^T)$,~
${\bm \Sigma}^r \equiv [ (1-\lambda) \bm{p}\bm{h} - (\lambda+1) \bm{h}\bm{p}]/2-\ell (\nabla \bm{h} + \nabla \bm{h}^T)$, and $\bm{h} = - \delta F / \delta {\bm p}$.
These are sufficient for our current purpose because we have already established in Section~\ref{subsub:smallq} that for
extensile systems, which is our main focus, the dominant failure modes are the twist bend modes which are always decoupled from
concentration fluctuations. The dispersion relation for these modes with these terms is:
\begin{eqnarray} \label{tbpolar}
    \omega^t_{\pm} ={1 \over 2}v_0 q\cos\phi - i{\mu_+ \over 2 \rho} q^2
    \pm \left(\sigma_a \over 2\rho\right)^{1/2} \left[A(0)\cos^2\phi q^2+iB(0)\left(1-{\gamma_a(1+\lambda)+2\ell\sigma_a \over v_0\mu_{-}}\right)\cos\phi q^3+\left(\bar{G}(\phi)+{2\gamma_a \ell \over \sigma_a}\right)q^4 \right]^{1/2}\nonumber\\
\end{eqnarray}
where $R \equiv \rho v_0^2/2 \sigma_a$,  $\mu_{\pm}={\mu} (1 \pm \beta)$, $\beta = \Gamma K \rho/\mu$, $B(\phi)= (v_0\mu_-/\sigma_a) \cos\phi$,  $G(\phi)=-(\mu_-^2/2 \rho \sigma_a) + (K/2\sigma_a)(1+\lambda\cos 2\phi)^2$, and $\bar{G}(\phi) = - (\mu_-^2/2 \rho \sigma_a) +  (K/2\sigma_a)(1+\lambda)^2\cos^2\phi$.
Comparing \eqref{tbpolar} to the dispersion relation for $\gamma_a, \ell = 0$ [Eq.~\eqref{eq:omvsq34} of the main text] 
\begin{eqnarray}
\label{tbwopolar}
\omega^t_{\pm}= {1 \over 2}v_0 q\cos\phi- i{\mu_+ \over 2 \rho} q^2 \pm \left(\sigma_a \over 2\rho\right)^{1/2} \left[A(0)\cos^2\phi q^2+ iB(0)\cos\phi q^3+\bar{G}(\phi)q^4 \right]^{1/2}. 
\end{eqnarray}
and working to order $q^2$ we see 
that the transition between $O(q)$ and $O(q^2)$ 
instabilities, which is determined by $A(0)$, remains unchanged and that the relative shifts of the coefficients of $q^2$ in the mode frequencies are of order
\begin{equation} 
\label{activecap}  
D_1 = \gamma_a / \mu v_0, \, \, D_2 = \ell \sigma_a/ \mu v_0
\end{equation} 
which resemble inverse capillary numbers given that $\gamma_a$ and $\ell \sigma_a$ have units of surface tension.
The length $\ell$ controls polar couplings of orientation to flow in the passive theory and as such should be related, in a microscopic theory, to a geometrical measure of polarity of the constituent particles, such as a fore-aft size difference. $\gamma_a$, with units of force per unit length, governs polar active contributions to the stress tensor. It is then reasonable to expect that $\gamma_a \sim \sigma_a \ell$, and thus that $D_1$ and $D_2$ are similar in magnitude. For particles of size $b$ the viscosity-based estimate $\sigma_a \sim \mu v_0 / b$ implies $D_1, D_2 \sim \ell/b$, while an inertia-based estimate $\sigma_a \sim \rho v_0^2$ yields $D_1, D_2 \sim \ell Re/b$ where $Re = \rho v_0 b / \mu$ is the Reynolds number at the particle scale. With either estimate, $D_1$ and $D_2$ should typically be small because $\ell/b$ is the ratio of a fore-aft size difference to an overall size. Fig.~\ref{fig:Rvsbetapolar} shows the results of linear stability analysis for nonzero $D_1 = D_2$. The $O(q)$-unstable, $O(q^2)$-unstable and linearly stable regimes exist over the entire range; for $D_1, D_2 \ll 1$ the linear stability analysis differs negligibly from that with $\gamma_a, \ell =0$; and the $O(q^2)$-unstable regime shrinks in extent as $D_1,D_2$ grow to values of order unity.
We therefore expect the qualitative features of the dynamical phase diagram as seen in our numerical studies to persist for nonzero $\gamma_a$ and $\ell$, but we have not carried out the corresponding direct numerical solutions of the hydrodynamic equations.

\begin{thebibliography}{108}
\bibitem [{con()}]{concdepfoot}%
  \BibitemOpen
  \href@noop {} {}\bibinfo {note} {We assume concentrations are high enough
  that {$F$} globally favours a nonzero magnitude for $\bm{p}$. We therefore
  ignore $c$-dependence in the polynomial part of {$F$}.}\BibitemShut {Stop}%
\bibitem [{GGb()}]{GGbarfoot}%
  \BibitemOpen
  \href@noop {} {}\bibinfo {note} {In molecular or colloidal systems $\mu /\rho
  \gg {K}/\mu$. Assuming this inequality holds for the systems of interest here
  as well, {$G$} and $\bar{G}$ are uniformly negative.}\BibitemShut {Stop}%
\bibitem [{foo()}]{foot:quad}%
  \BibitemOpen
  \href@noop {} {}\bibinfo {note} {Even for slow swimmers, one could imagine
  inertia, in the form of {$R$}, overcoming the force-dipole instability in the
  case of a swimmer with a small dipole (and hence a small $\sigma_0$),
  propelled mainly by its quadrupole force moment.}\BibitemShut {Stop}%
\bibitem [{FSS()}]{FSS}%
  \BibitemOpen
  \href@noop {} {}\bibinfo {note} {Such analyses are most useful in the context
  of an exchange of dominance between system size $L$ and correlation length
  $\xi$ as one goes from $L<\xi$ to $L>\xi$, with no other scale playing a
  role. This requires studies at several values of $L$ substantially larger
  than a short-distance cutoff below which lie the fast modes which contribute
  the noise actuating the slow modes. For our system, this cutoff scale is the
  wavelength of the fastest-growing modes of the unstable linear theory. We
  defer the required exceedingly large-scale numerical studies to later
  work}\BibitemShut {NoStop}%
\end{thebibliography}
\providecommand{\noopsort}[1]{}\providecommand{\singleletter}[1]{#1}%
%
\end{document}